\documentclass[12pt]{article}
\usepackage{amsmath,amsthm}
\usepackage{eufrak}
\usepackage[pdftex]{graphicx}

\newcommand{\ar}{\renewcommand{\arraystretch}{1.3}} 
\DeclareMathAlphabet{\bb}{U}{msb}{m}{n}  \gdef\dZ{\bb
Z}    \gdef\R{\bb R}

 \DeclareMathOperator{\spin}{{\bf
Spin}}

\DeclareMathOperator{\Sym}{Sym}

\DeclareMathOperator{\SO}{SO}\DeclareMathOperator{\SU}{SU}
 \DeclareMathOperator{\GO}{O}

\newcommand{\cH}{{\cal H}}

\newcommand{\sA}{{\sf A}}

\newcommand{\sL}{{\sf L}}

\newcommand{\bsH}{{\boldsymbol{\sf H}}}

\newcommand{\bsT}{{\boldsymbol{\sf T}}}

\newcommand{\bsL}{{\boldsymbol{\sf L}}}
\newcommand{\bsK}{{\boldsymbol{\sf K}}}
\newcommand{\bsJ}{{\boldsymbol{\sf J}}}
\newcommand{\bsS}{{\boldsymbol{\sf S}}}
\newcommand{\bsQ}{{\boldsymbol{\sf Q}}}
\newcommand{\bsP}{{\boldsymbol{\sf P}}}

\newcommand{\fP}{\mathfrak{P}}
\newcommand{\fK}{\mathfrak{K}}

\newcommand{\cl}{C\kern -0.2em \ell}

\newcommand{\ld}{\left[}
\newcommand{\rd}{\right]}
\newcommand{\lf}{\left\{}
\newcommand{\rf}{\right\}}
\begin{document}
\title{The Periodic Table and the Group $\SO(4,4)$:\\ II. Double $\SO(4,2)$-tower}
\author{V.~V. Varlamov\thanks{Siberian State Industrial University,
Kirova 42, Novokuznetsk 654007, Russia, e-mail: vadim.varlamov@mail.ru}}
\date{}
\maketitle
\begin{abstract}
A group-theoretic interpretation of the periodic system of elements is given within the framework of the weight diagram of the Lie algebra $\mathfrak{so}(4,4)$ of the fourth rank, where the four quantum numbers $n$, $l$, $m$, $s$ correspond to the eigenvalues (weights) of the Cartan generators of the maximal Abelian subalgebra (the maximal torus of the group $\SO(4,4)$). It is shown that the root system of the algebra $\mathfrak{so}(4,4)$ forms a regular four-dimensional self-dual polyhedron (24-cell). The action of the fourth Cartan generator associated with spin leads to a splitting of the Cartan-Weyl basis of the algebra $\mathfrak{so}(4,4)$ into two structurally identical bases, each of which is isomorphic to the Yao basis of the subalgebra $\mathfrak{so}(4,2)$ (the Lie algebra of the conformal group). At this point, a four-dimensional 24-cell is projected onto two three-dimensional cuboctahedra, each of which defines the root system of the subalgebra $\mathfrak{so}(4,2)$. This splitting physically corresponds to spin doubling (two-valuedness). The structure of the energy levels of a periodic system is studied, the states of which (chemical elements) are represented as nodes of the weight diagram of the group algebra $\mathfrak{so}(4,4)$. The structure of the double $\SO(4,2)$-towers of Mendeleev, Seaborg, and 10-periodic extension is examined in detail. The period doubling associated with the sequence of period lengths 2, 8, 8, 18, 18, 32, 32, $\ldots$ of the periodic system of elements is explained by the action of the fourth Cartan generator. It is shown that antimatter (Mendeleev anti-table consisting of antihydrogen, antihelium, antilitium, $\ldots$) is naturally included in the general group-theoretic scheme of description of the periodic table.
\end{abstract}

\section{Introduction\label{sec1}}
As is well known, the Bohr's Aufbau scheme \cite{Bohr} is historically the first attempt to provide a theoretical explanation of the structure of the periodic table of chemical elements. This scheme is based on the Bohr-Sommerfeld model of the atom, according to which the atom has an internal structure similar to the solar system in miniature (Rutherford's planetary model). A kind of symbiosis of the classical Kepler problem and quantum postulates\footnote{This leads to a logical inconsistency, since the model is neither classical nor quantum: in the system of two equations underlying it, one is the classical equation of electron motion, and the other is the quantum equation of orbit quantization.} allowed Bohr to derive the Rydberg formula and calculate the energy spectrum of the hydrogen atom (the main spectral series of Lyman, Balmer and Paschen). However, apart from the hydrogen spectrum\footnote{The explanation of the existence of a thin and hyperfine structure of spectral lines of hydrogen goes beyond the description of the Bohr model, as well as the description of the Zeeman effect and anomalous ultraviolet Pickering-Fowler series.}, the model proved unable to describe the energy levels of multi-electron atoms (already starting with the helium atom).

Bohr's model assumes that electrons have known orbits and locations - two things that cannot be measured simultaneously according to the Heisenberg uncertainty principle. An electron, as a quantum object, does not have the concept of a classical trajectory. Heisenberg writes: ``$\ldots$ the idea of an electronic orbit, associated with the idea of a discrete stationary state, was practically discarded along the way. The concept of discrete stationary states, however, has remained alive. The concept was necessary. It had its basis in observational data. On the contrary, the electronic orbit could not be reconciled with the observations, so it was abandoned, and only the matrices for coordinates remained'' \cite[c.~97]{Heisen}. Discrete stationary states corresponding to the eigenvalues of the energy operator are what remains of the electronic orbits of the planetary model. However, the concept of an electron orbit turned out to be extremely tenacious, transforming into the concept of an atomic orbital, i.e. into the idea that an electron in an atom looks like some kind of electron cloud, the various shapes of which are determined depending on the quantum numbers $n$, $l$ and $m$. An important point should be noted here: all atomic orbitals are described by the three quantum numbers $n$, $l$ and $m$, the fourth quantum number $s$ (spin) does not participate in the definition of atomic orbitals. As for the real existence of atomic orbitals, it makes no sense to talk about their real existence. This is a mathematical abstraction, the issue has been closed in principle since the time of Heisenberg\footnote{However, in 1999 there appeared an article \cite{Orbit} in the prestigious journal ``Nature'', which talked about supposedly experimentally observed atomic orbitals. The controversy arose \cite{Sherri,Orbit2}, as a result of which it turned out that the authors of the ``discovery'' actually saw electron density, and not atomic orbitals.}.

The structure of the periodic table, based on the Bohr model (Aufbau scheme), assumes that the arrangement of elements in the system as their atomic numbers increase is uniquely determined by the individual features of the electronic structure of atoms described in the framework of the one-electron approximation (Hartree method), and directly reflects the energy sequence of atomic orbitals $s$-, $p$-, $d$-, $f$-shells populated by electrons as their total number increases as the charge of the atomic nucleus increases in accordance with the principle of minimum energy. However, this is possible only in the simplest version of the Hartree approximation, but in the version of the Hartree-Fock approximation, the total energy of an atom is not equal to the sum of the orbital energies, and the electronic configuration of an atom is determined by the minimum of its total energy. The Bohr model does not explain the periodicity, but only approximates it within the framework of the one-electron Hartree approximation. Moreover, Bohr derived electronic configurations not from quantum theory, but from the known chemical and spectroscopic properties of the elements.

It is easy to see that the Aufbau scheme implements the classical reductionist principle, according to which \textit{whole} is defined by \textit{parts}, i.e. the whole, understood as \textit{aggregate}, is derived from its parts. The \textit{Holistic antithesis} to the Aufbau scheme is the Rumer-Fet-Barut model \cite{RF71,Bar72}. A characteristic feature of this model is the representation of the periodic system of elements as a \textit{single quantum system}, the states of which (chemical elements) are connected to each other under the action of a symmetry group. In this case, the symmetry group is usually understood as tensor extensions of a conformal group: $G_F=\SO(4,2)\otimes\SU(2)\otimes\SU(2)^\prime$ (Fet group), $G_O=\GO(4,2)\otimes\SU(2)_S\otimes\SU(2)_T$ (Ostrovsky group). The group approach, in general, represents the most suitable mathematical structure for describing the phenomenon of periodicity, i.e., repeatability (cyclicity), which, as is known, is the main characteristic feature of a system of chemical elements.

This article is a continuation of the work  \cite{Var2401,Var2402,Var2501,Var2502}, which solves the problem of constructing a periodic system of chemical elements within the framework of the weight diagram of the Lie algebra of the fourth rank of the rotation group $\SO(4,4)$ of an eight-dimensional pseudo-Euclidean space $\R^{4,4}$ of a neutral signature. Historically, graphical visualization of the periodic law has been very important. In addition to the initial 2-dimensional tabular representation (the Mendeleev and Meyer tables), there are many other different forms: spiral, helical and pyramidal models of the periodic table, as well as many exotic forms (for more details, see \cite{Spronson,Mazurs,Scerri}. The main feature of these models is the construction of a system of chemical elements depending on the increase in atomic weight. However, quantum mechanics has led to the understanding that the main structural characteristic of the periodic law is not a linear increase in atomic weight, but a structure determined by the order of quantum numbers. Polygonfl\"{a}che by G. Haenzel \cite{Haenzel} is historically the first model of its kind, where the structure of a periodic system is determined by the order of quantum numbers forming systems of concentric polygons on a plane. V. Finke in the article \cite{Finke} presents Haenzel's Polygonfl\"{a}che in three-dimensional space, linking it with the Madelung numbering. On the other hand, in the Rumer-Fet-Barut approach, the structure of the "left-sided" Janet table is associated with the quantum numbers of the conformal group $\SO(4,2)$ and its tensor extension $G_F$ \cite{RF71,Fet}, see also \cite{Var1801,Var1802,Var1901}. As is known, the Janet table is a two-dimensional graphical representation of the Madelung rule. Thus, by means of Madelung numbering, there is a correspondence between the Haenzel-Finke model and the Rumer-Fet-Barut group-theoretic approach. However, a common disadvantage of Haenzel's Polygonfl\"{a}che and Finke's three-dimensional system is the artificial representation of the fourth quantum number $s$ (spin) as two points on transversals. In turn, in the group-theoretic approach, the first three quantum numbers $n$, $l$, and $m$ correspond to the eigenvalues of Cartan generators forming the maximal Abelian subalgebra of the Lie algebra $\mathfrak{so}(4,2)$ of the conformal group. The algebra $\mathfrak{so}(4,2)$ has the third rank. To include the fourth quantum number $s$ on an equal basis with $n$, $l$, $m$, i.e. as an eigenvalue of the Cartan generator, a transition to the Lie algebra of the fourth rank is required. It is shown in \cite{Var2501,Var2502} that such an algebra is $\mathfrak{so}(4,4)$ -- the Lie algebra of the group $\SO(4,4)$. The corresponding spin generator $\bsL_{78}$ commutes with all generators of the subalgebra $\mathfrak{so}(4,2)$, which leads to splitting (doubling) of the Cartan-Weyl basis of the algebra $\mathfrak{so}(4,4)$. As a consequence, the four-dimensional weight diagram of the algebra $\mathfrak{so}(4,4)$ can be represented by two three-dimensional projections, each of which is isomorphic to the weight diagram of the subalgebra $\mathfrak{so}(4,2) $ \cite{Var2501,Var2502}. It should be noted that the group $\SO(4,4)$ has a higher symmetry than the tensor extensions $G_F$ and $G_O$: the Lie algebra $\mathfrak{so}(4,4)$ contains 28 independent generators, while the Lie algebra $\mathfrak{so}(4,2)\otimes\mathfrak{su}(2)\otimes\mathfrak{su}(2)^\prime$ of the groups $G_F$ and $G_O$ has 21 independent generators. The main disadvantage of the Ostrovsky-Fet scheme of the group-theoretic description of period doubling is the artificial nature of the introduction of the fifth quantum number, which has no real analogue, since all states (elements) of the periodic system are described by the four quantum numbers $(n,l,m.s)$. Moreover, the introduction of the fifth quantum number $s^\prime$ leads to a change in the Madelung rule (the Fet rule \cite{Fet}). It is shown in section 7 that the fourth Cartan generator $\bsL_{78}$ (spin generator), due to the higher symmetry of the group $\SO(4,4)$, combines (``absorbs'') the functions of the Cartan generators $\boldsymbol{\tau}_3$ and $\boldsymbol{\tau}^\prime_3$ of the Ostrovsky-Fet scheme, which avoids the introduction of an additional (non-physical) fifth quantum number $s^\prime$. In section 3 of this article, two three-dimensional projections of the weight diagram of the algebra $\mathfrak{so}(4,4)$ are combined into a double $\SO(4,2)$-tower, each node of which (the finite-dimensional representation of the group $\SO(4,2)$) is associated with the corresponding element of the periodic table according to the Madelung numbering. The sweep of the double $\SO(4,2)$-tower onto the Janet table and the correspondence with the Rumer-Fet-Barut model is established through the Madelung basis. It is shown in section 4 that the main energy levels (floors $\SO(4,2)$-towers), determined by the eigenvalues of the radial generator $\bsL_{56}$, have the structure of weight $(j,j)$-diagrams of the subgroup $\SO(4)$, where $j$ is an integer. The $\SO(4,2)$-towers have a similar structure of energy levels, defining the 8-periodic and 10-periodic extensions of the periodic system (see sections 5 and 6). It is shown in the section 8 that antimatter (the Mendeleev anti-table consisting of antielements: antihydrogen, antihelium, antilitium, and $\ldots$) is naturally included in the general group-theoretic scheme for describing the periodic table.

\section{The Madelung Rule}
As is known, as the charge of the atomic nucleus $Z$ increases in the periodic table, the filling of orbitals in a neutral atom is set in a certain sequence, which is traditionally called the Aufbau scheme \cite{Bohr}. Obviously, the Aufbau circuit is determined by a sequence of single-electron levels $\varepsilon_{nl}$ on the energy scale. This sequence depends on the properties of the effective one-electron central field (Coulomb field with potential $-Z/r$). The lengths of the periods in the hydrogen-like Aufbau scheme correspond to hydrogen $n$ shells: $2n^2=2,8,18,32,50,\ldots$. A comparison with detailed numerical quantum calculations for multielectronic systems and empirical data shows that for \textit{highly ionized} atoms, the $(n,l)$ ordering really works \cite{Ost96}.

However, the structure of \textit{neutral atoms} is of paramount importance when it comes to the periodic table. The $(n,l)$ rule is not satisfied here. This means that the effective potential is very different from the Coulomb potential, which leads to a significant restructuring of the spectrum. There is an overlap between groups of energy levels with different principal quantum numbers $n$. The $n$-grouping of levels disappears and new types of regularities appear \cite{DO72,Ost96,KK05}.

Empirically, it has been observed that the periodic table is well described by the Madelung rule ($n+l,n$): orbitals are filled in ascending order of the sum $N\equiv n+l$, and for a fixed $N$ - in ascending order of $n$. The advantage of the ($n+l,n$) scheme in comparison with the rule $(n,l)$ is especially clearly seen from the graphical construction given by Demkov and Ostrovsky \cite{DO72}.

A more detailed comparison of the sequences of single-electron levels in both circuits further demonstrates the advantage of the Madelung rule:\\
\textbf{Aufbau scheme $(n,l)$}:
\begin{equation}
\overbrace{\underbrace{1s}_{\dim=2}}^{n=1}\ll\overbrace{\underbrace{2s<2p}_{\dim=8}}^{n=2}
\ll\overbrace{\underbrace{3s<3p<3d}_{\dim=18}}^{n=3}\ll\overbrace{\underbrace{4s<4p<4d<4f}_{\dim=32}}^{n=4}
\ll\overbrace{\underbrace{5s<5p<5d<5f<5g}_{\dim=50}}^{n=5}\ll\ldots.
\end{equation}
\textbf{The Madelung rule $(n+l,n)$}:
\begin{multline}
\overbrace{\underbrace{1s}_{\dim=2}}^{n+l=1}\ll\overbrace{\underbrace{2s}_{\dim=2}}^{n+l=2}<
\overbrace{\underbrace{2p\ll 3s}_{\dim=8}}^{n+l=3}<\overbrace{\underbrace{3p\ll 4s}_{\dim=8}}^{n+l=4}\ll
\overbrace{\underbrace{3d<4p\ll 5s}_{\dim=18}}^{n+l=5}<\\
<\overbrace{\underbrace{4d<5p\ll 6s}_{\dim=18}}^{n+l=6}<\overbrace{\underbrace{4f<5d<6d\ll 7s}_{\dim=32}}^{n+l=7}<
\overbrace{\underbrace{5f<6d<7p\ll 8s}_{\dim=32}}^{n+l=8}<\ldots.
\end{multline}
The initial parts match in both cases, but later the differences become significant. It is clearly seen that in both cases the same dimensions appear for $n$- and $(n+l)$-shells, but for the scheme $(n+l,n)$, each of them occurs twice.

\begin{figure}[ht]
\centering
\includegraphics[width=\textwidth]{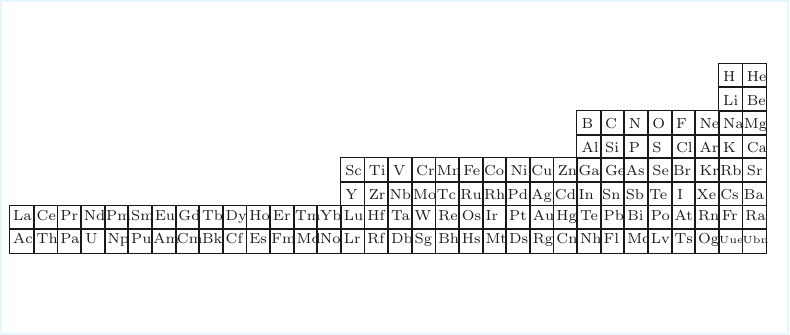}\\
\caption{Janet left-step table of chemical elements (1929).}
\label{pic1}
\end{figure}

The Janet table, shown in Figure \ref{pic1}, is directly related to the Madelung rule.
To this form of the table \cite{Janet} Janet came by presenting the periodic table as a collection of four nested cylinders. On the surface of each cylinder, chemical elements are arranged in a helical line (by analogy with the Telluric screw of Chancourtois in 1852). Next, by projecting the cylinders onto a plane, Janet obtains a spiral\footnote{It should be noted that the Janet spiral was not the first of a wide variety of spiral models of the periodic table.}. The unfolding of this spiral leads to the left-step Janet table.

\begin{figure}[p] %
\centering
\includegraphics[width=\textwidth]{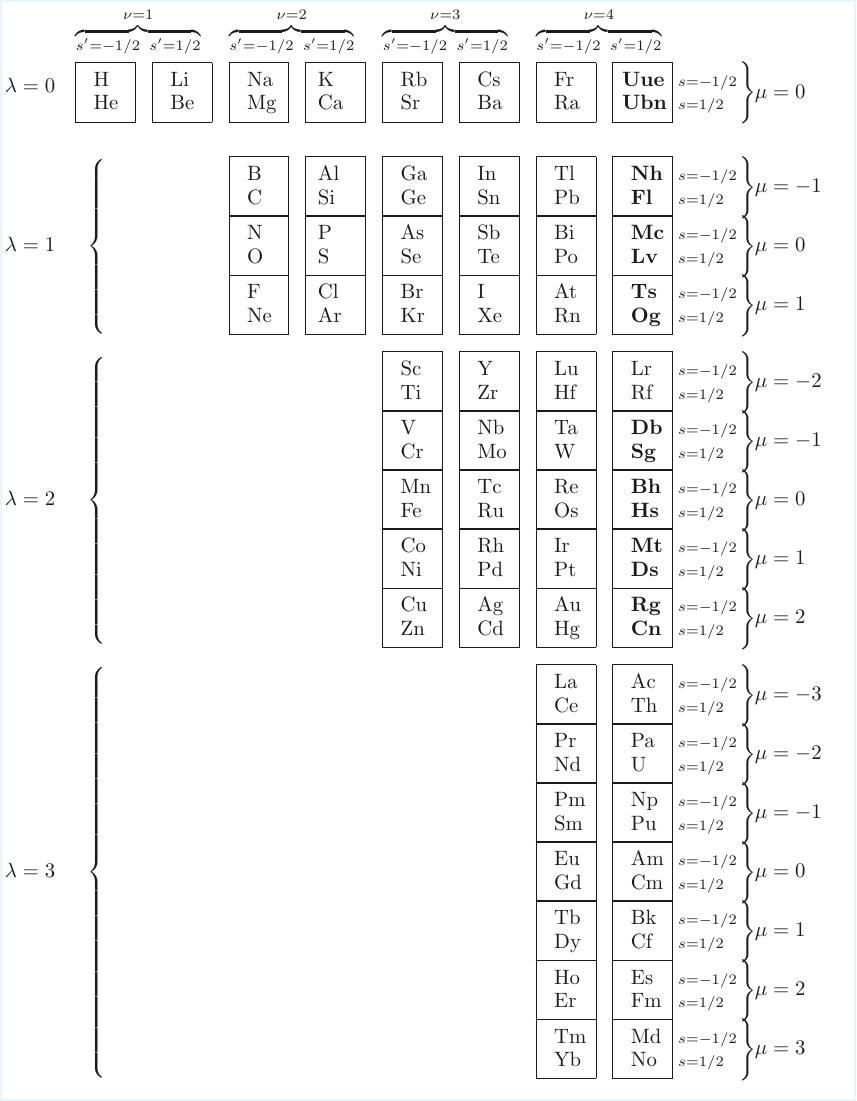}\\
\caption{The periodic table in the Janet-like form of the basic representation  $F^+_{ss^\prime}$ of the Fet group $\SO(2,4)\otimes\SU(2)\otimes\SU(2)^\prime$. Quantum numbers $(\nu,\lambda,\mu,s,s^\prime)$ correspond to the first Fet basis \cite{Fet}.}
\label{pic_2}
\end{figure}

We see the same Janet table in the pioneering work of Yu.B. Rumer and A.I. Fet \cite{RF71}, devoted to the group-theoretic description of the periodic table (see also \cite{Fet,Var1801}). Only now this table is rotated by $90^\circ$ and is accompanied by the notation of the corresponding quantum numbers of the conformal group (see Figure \ref{pic_2}).

Rumer and Fet do not mention the Janet table anywhere, apparently it was unknown to them, and they came to this form of the periodic table independently as a result of studying the group properties of the periodic table. The 8-periodic extension (Seaborg table) of the periodic system (see Figures 1--2 in \cite{Var1802}), as well as the 10-periodic extension (see Figure 1 in \cite{Var1901}) can be represented in the same form (Janet left-step table).

Period doubling means that the entire variety of chemical elements naturally splits into two sets with the sum of $n+l$, even or odd, where $n$ and $l$ are the main and orbital quantum numbers, respectively. As a result, elements from the same (even or odd) set are chemically more similar than elements from different sets \cite{Var2401}.

Within the framework of the group-theoretic description \cite{Bar72,Fet}, chemical elements are represented by various states
\begin{equation}\label{Ket1}
|Z\rangle=|n,l,m,s\rangle
\end{equation}
of a single quantum system, where the ket-vector of the form (\ref{Ket1}) depends on four quantum numbers\footnote{The group-theoretic meaning of quantum numbers will be discussed below in paragraph \,3.} of a hydrogen-like system and forms the basis of an infinite-dimensional Hilbert space. In Table 1, the ket-vectors (\ref{Ket1}) are arranged in ascending order of atomic number $Z$ according to the Madelung numbering. The first column shows the atomic number of the element. The second column contains the element symbol. The third column contains the ket-vector corresponding to the atomic number $Z$. The fourth column contains the sum $n+l$, the even '$+$' and the odd '$-$'. The fifth column shows the electronic configuration (according to the Aufbau scheme).
\newpage
\begin{center}
{\textbf{Table\,1.} The Madelung basis}
\end{center}
\begin{center}
\vspace{-0.2cm}
{\renewcommand{\arraystretch}{1.0}
\begin{tabular}{|c|l|l|c|l|}\hline
$Z$ & Element     & Vector $|n,l,m,s\rangle$ & $n+l$ & Configuration  \\ \hline\hline
1 & \textbf{H}& $|1,0,0,-1/2\rangle$ & $-$ & $1s^1$ \\
2 & \textbf{He}& $|1,0,0,+1/2\rangle$ & $-$ & $1s^2$ \\
3 & \textbf{Li}& $|2,0,0,-1/2\rangle$ & $+$ & [\textbf{He}]$2s^1$ \\
4 & \textbf{Be}& $|2,0,0,+1/2\rangle$ & $+$ & [\textbf{He}]$2s^2$ \\
5 & \textbf{B}& $|2,1,-1,-1/2\rangle$ & $-$ & [\textbf{He}]$2s^22p^1$ \\
6 & \textbf{C}& $|2,1,0,-1/2\rangle$ & $-$ & [\textbf{He}]$2s^22p^2$ \\
7 & \textbf{N}& $|2,1,1,-1/2\rangle$ & $-$ & [\textbf{He}]$2s^22p^3$ \\
8 & \textbf{O}& $|2,1,-1,+1/2\rangle$ & $-$ & [\textbf{He}]$2s^22p^4$ \\
9 & \textbf{F}& $|2,1,0,+1/2\rangle$ & $-$ & [\textbf{He}]$2s^22p^5$ \\
10 & \textbf{Ne}& $|2,1,1,+1/2\rangle$ & $-$ & [\textbf{He}]$2s^22p^6$ \\
11 & \textbf{Na}& $|3,0,0,-1/2\rangle$ & $-$ & [\textbf{Ne}]$3s^1$ \\
12 & \textbf{Mg}& $|3,0,0,+1/2\rangle$ & $-$ & [\textbf{Ne}]$3s^2$ \\
13 & \textbf{Al}& $|3,1,-1,-1/2\rangle$ & $+$ & [\textbf{Ne}]$3s^23p^1$ \\
14 & \textbf{Si}& $|3,1,0,-1/2\rangle$ & $+$ & [\textbf{Ne}]$3s^23p^2$ \\
15 & \textbf{P}& $|3,1,1,-1/2\rangle$ & $+$ & [\textbf{Ne}]$3s^23p^3$ \\
16 & \textbf{S}& $|3,1,-1,+1/2\rangle$ & $+$ & [\textbf{Ne}]$3s^23p^4$ \\
17 & \textbf{Cl}& $|3,1,-1,+1/2\rangle$ & $+$ & [\textbf{Ne}]$3s^23p^5$ \\
18 & \textbf{Ar}& $|3,1,1,+1/2\rangle$ & $+$ & [\textbf{Ne}]$3s^23p^6$ \\
19 & \textbf{K}& $|4,0,0,-1/2\rangle$ & $+$ & [\textbf{Ar}]$4s^1$ \\
20 & \textbf{Ca}& $|4,0,0,+1/2\rangle$ & $+$ & [\textbf{Ar}]$4s^2$ \\
21 & \textbf{Sc}& $|3,2,-2,-1/2\rangle$ & $-$ & [\textbf{Ar}]$3d^14s^2$ \\
22 & \textbf{Ti}& $|3,2,-1,-1/2\rangle$ & $-$ & [\textbf{Ar}]$3d^24s^2$ \\
23 & \textbf{V}& $|3,2,0,-1/2\rangle$ & $-$ & [\textbf{Ar}]$3d^34s^2$ \\
24 & \textbf{Cr}& $|3,2,1,-1/2\rangle$ & $-$ & [\textbf{Ar}]$3d^54s^1$ \\
25 & \textbf{Mn}& $|3,2,2,-1/2\rangle$ & $-$ & [\textbf{Ar}]$3d^54s^2$ \\
26 & \textbf{Fe}& $|3,2,-2,+1/2\rangle$ & $-$ & [\textbf{Ar}]$3d^64s^2$ \\
27 & \textbf{Co}& $|3,2,-1,+1/2\rangle$ & $-$ & [\textbf{Ar}]$3d^74s^2$ \\
28 & \textbf{Ni}& $|3,2,0,+1/2\rangle$ & $-$ & [\textbf{Ar}]$3d^84s^2$ \\
29 & \textbf{Cu}& $|3,2,1,+1/2\rangle$ & $-$ & [\textbf{Ar}]$3d^{10}4s^1$ \\
30 & \textbf{Zn}& $|3,2,2,+1/2\rangle$ & $-$ & [\textbf{Ar}]$3d^{10}4s^2$ \\
31 & \textbf{Ga}& $|4,1,-1,-1/2\rangle$ & $-$ & [\textbf{Ar}]$3d^{10}4s^24p^1$ \\
32 & \textbf{Ge}& $|4,1,0,-1/2\rangle$ & $-$ & [\textbf{Ar}]$3d^{10}4s^24p^2$ \\
33 & \textbf{As}& $|4,1,1,-1/2\rangle$ & $-$ & [\textbf{Ar}]$3d^{10}4s^24p^3$ \\
34 & \textbf{Se}& $|4,1,-1,+1/2\rangle$ & $-$ & [\textbf{Ar}]$3d^{10}4s^24p^4$ \\
35 & \textbf{Br}& $|4,1,0,+1/2\rangle$ & $-$ & [\textbf{Ar}]$3d^{10}4s^24p^5$ \\
36 & \textbf{Kr}& $|4,1,1,+1/2\rangle$ & $-$ & [\textbf{Ar}]$3d^{10}4s^24p^6$ \\
37 & \textbf{Rb}& $|5,0,0,-1/2\rangle$ & $-$ & [\textbf{Kr}]$5s^1$ \\
38 & \textbf{Sr}& $|5,0,0,+1/2\rangle$ & $-$ & [\textbf{Kr}]$5s^2$ \\
39 & \textbf{Y}& $|4,2,-2,-1/2\rangle$ & $+$ & [\textbf{Kr}]$4d^15s^2$ \\
40 & \textbf{Zr}& $|4,2,-1,-1/2\rangle$ & $+$ & [\textbf{Kr}]$4d^25s^2$ \\
41 & \textbf{Nb}& $|4,2,0,-1/2\rangle$ & $+$ & [\textbf{Kr}]$4d^45s^1$ \\
42 & \textbf{Mo}& $|4,2,1,-1/2\rangle$ & $+$ & [\textbf{Kr}]$4d^55s^1$ \\
43 & \textbf{Tc}& $|4,2,2,-1/2\rangle$ & $+$ & [\textbf{Kr}]$4d^55s^2$ \\
44 & \textbf{Ru}& $|4,2,-2,+1/2\rangle$ & $+$ & [\textbf{Kr}]$4d^75s^1$ \\
45 & \textbf{Rh}& $|4,2,-1,+1/2\rangle$ & $+$ & [\textbf{Kr}]$4d^85s^1$ \\
\hline
\end{tabular}
}
\end{center}
\begin{center}
{\renewcommand{\arraystretch}{1.0}
\begin{tabular}{|c|l|l|c|l|}\hline
$Z$ & Element     & Vector $|n,l,m,s\rangle$ & $n+l$ & Configuration  \\ \hline\hline
46 & \textbf{Pd}& $|4,2,0,+1/2\rangle$ & $+$ & [\textbf{Kr}]$4d^{10}$ \\
47 & \textbf{Pd}& $|4,2,1,+1/2\rangle$ & $+$ & [\textbf{Kr}]$4d^{10}5s^1$ \\
48 & \textbf{Cd}& $|4,2,2,+1/2\rangle$ & $+$ & [\textbf{Kr}]$4d^{10}5s^2$ \\
49 & \textbf{In}& $|5,1,-1,-1/2\rangle$ & $+$ & [\textbf{Kr}]$4d^{10}5s^25p^1$ \\
50 & \textbf{Sn}& $|5,1,0,-1/2\rangle$ & $+$ & [\textbf{Kr}]$4d^{10}5s^25p^2$ \\
51 & \textbf{Sb}& $|5,1,1,-1/2\rangle$ & $+$ & [\textbf{Kr}]$4d^{10}5s^25p^3$ \\
52 & \textbf{Te}& $|5,1,-1,+1/2\rangle$ & $+$ & [\textbf{Kr}]$4d^{10}5s^25p^4$ \\
53 & \textbf{I}& $|5,1,0,+1/2\rangle$ & $+$ & [\textbf{Kr}]$4d^{10}5s^25p^5$ \\
54 & \textbf{Xe}& $|5,1,1,+1/2\rangle$ & $+$ & [\textbf{Kr}]$4d^{10}5s^25p^6$ \\
55 & \textbf{Cs}& $|6,0,0,-1/2\rangle$ & $+$ & [\textbf{Xe}]$6s^1$ \\
56 & \textbf{Ba}& $|6,0,0,+1/2\rangle$ & $+$ & [\textbf{Xe}]$6s^2$ \\
57 & \textbf{La}& $|4,3,-3,-1/2\rangle$ & $-$ & [\textbf{Xe}]$5d^16s^2$ \\
58 & \textbf{Ce}& $|4,3,-2,-1/2\rangle$ & $-$ & [\textbf{Xe}]$4f^15d^16s^2$ \\
59 & \textbf{Pr}& $|4,3,-1,-1/2\rangle$ & $-$ & [\textbf{Xe}]$4f^36s^2$ \\
60 & \textbf{Nd}& $|4,3,0,-1/2\rangle$ & $-$ & [\textbf{Xe}]$4f^46s^2$ \\
61 & \textbf{Pm}& $|4,3,1,-1/2\rangle$ & $-$ & [\textbf{Xe}]$4f^56s^2$ \\
62 & \textbf{Sm}& $|4,3,2,-1/2\rangle$ & $-$ & [\textbf{Xe}]$4f^66s^2$ \\
63 & \textbf{Eu}& $|4,3,3,-1/2\rangle$ & $-$ & [\textbf{Xe}]$4f^76s^2$ \\
64 & \textbf{Gd}& $|4,3,-3,+1/2\rangle$ & $-$ & [\textbf{Xe}]$4f^75d^16s^2$ \\
65 & \textbf{Tb}& $|4,3,-2,+1/2\rangle$ & $-$ & [\textbf{Xe}]$4f^96s^2$ \\
66 & \textbf{Dy}& $|4,3,-1,+1/2\rangle$ & $-$ & [\textbf{Xe}]$4f^{10}6s^2$ \\
67 & \textbf{Ho}& $|4,3,0,+1/2\rangle$ & $-$ & [\textbf{Xe}]$4f^{11}6s^2$ \\
68 & \textbf{Er}& $|4,3,1,+1/2\rangle$ & $-$ & [\textbf{Xe}]$4f^{12}6s^2$ \\
69 & \textbf{Tm}& $|4,3,2,+1/2\rangle$ & $-$ & [\textbf{Xe}]$4f^{13}6s^2$ \\
70 & \textbf{Yb}& $|4,3,3,+1/2\rangle$ & $-$ & [\textbf{Xe}]$4f^{14}6s^2$ \\
71 & \textbf{Lu}& $|5,2,-2,-1/2\rangle$ & $-$ & [\textbf{Xe}]$4f^{14}5d^16s^2$ \\
72 & \textbf{Hf}& $|5,2,-1,-1/2\rangle$ & $-$ & [\textbf{Xe}]$4f^{14}5d^26s^2$ \\
73 & \textbf{Ta}& $|5,2,0,-1/2\rangle$ & $-$ & [\textbf{Xe}]$4f^{14}5d^36s^2$ \\
74 & \textbf{W}& $|5,2,1,-1/2\rangle$ & $-$ & [\textbf{Xe}]$4f^{14}5d^46s^2$ \\
75 & \textbf{Re}& $|5,2,2,-1/2\rangle$ & $-$ & [\textbf{Xe}]$4f^{14}5d^56s^2$ \\
76 & \textbf{Os}& $|5,2,-2,+1/2\rangle$ & $-$ & [\textbf{Xe}]$4f^{14}5d^66s^2$ \\
77 & \textbf{Ir}& $|5,2,-1,+1/2\rangle$ & $-$ & [\textbf{Xe}]$4f^{14}5d^76s^2$ \\
78 & \textbf{Pt}& $|5,2,0,+1/2\rangle$ & $-$ & [\textbf{Xe}]$4f^{14}5d^96s^1$ \\
79 & \textbf{Au}& $|5,2,1,+1/2\rangle$ & $-$ & [\textbf{Xe}]$4f^{14}5d^{10}6s^1$ \\
80 & \textbf{Hg}& $|5,2,2,+1/2\rangle$ & $-$ & [\textbf{Xe}]$4f^{14}5d^{10}6s^2$ \\
81 & \textbf{Tl}& $|6,1,-1,-1/2\rangle$ & $-$ & [\textbf{Xe}]$4f^{14}5d^{10}6s^26p^1$ \\
82 & \textbf{Pb}& $|6,1,0,-1/2\rangle$ & $-$ & [\textbf{Xe}]$4f^{14}5d^{10}6s^26p^2$ \\
83 & \textbf{Bi}& $|6,1,1,-1/2\rangle$ & $-$ & [\textbf{Xe}]$4f^{14}5d^{10}6s^26p^3$ \\
84 & \textbf{Po}& $|6,1,-1,+1/2\rangle$ & $-$ & [\textbf{Xe}]$4f^{14}5d^{10}6s^26p^4$ \\
85 & \textbf{At}& $|6,1,0,+1/2\rangle$ & $-$ & [\textbf{Xe}]$4f^{14}5d^{10}6s^26p^5$ \\
86 & \textbf{Rn}& $|6,1,1,+1/2\rangle$ & $-$ & [\textbf{Xe}]$4f^{14}5d^{10}6s^26p^6$ \\
87 & \textbf{Fr}& $|7,0,0,-1/2\rangle$ & $-$ & [\textbf{Rn}]$7s^1$ \\
88 & \textbf{Ra}& $|7,0,0,+1/2\rangle$ & $-$ & [\textbf{Rn}]$7s^2$ \\
89 & \textbf{Ac}& $|5,3,-3,-1/2\rangle$ & $+$ & [\textbf{Rn}]$6d^17s^2$ \\
90 & \textbf{Th}& $|5,3,-2,-1/2\rangle$ & $+$ & [\textbf{Rn}]$6d^27s^2$ \\
91 & \textbf{Pa}& $|5,3,-1,-1/2\rangle$ & $+$ & [\textbf{Rn}]$5f^26d^27s^2$ \\
\hline
\end{tabular}
}
\end{center}
\begin{center}
{\renewcommand{\arraystretch}{1.0}
\begin{tabular}{|c|l|l|c|l|}\hline
$Z$ & Element     & Vector $|n,l,m,s\rangle$ & $n+l$ & Configuration  \\ \hline\hline
92 & \textbf{U}& $|5,3,0,-1/2\rangle$ & $+$ & [\textbf{Rn}]$5f^36d^27s^2$ \\
93 & \textbf{Np}& $|5,3,1,-1/2\rangle$ & $+$ & [\textbf{Rn}]$5f^46d^27s^2$ \\
94 & \textbf{Np}& $|5,3,2,-1/2\rangle$ & $+$ & [\textbf{Rn}]$5f^67s^2$ \\
95 & \textbf{Am}& $|5,3,3,-1/2\rangle$ & $+$ & [\textbf{Rn}]$5f^77s^2$ \\
96 & \textbf{Cm}& $|5,3,-3,+1/2\rangle$ & $+$ & [\textbf{Rn}]$5f^76d^17s^2$ \\
97 & \textbf{Bk}& $|5,3,-2,+1/2\rangle$ & $+$ & [\textbf{Rn}]$5f^97s^2$ \\
98 & \textbf{Cf}& $|5,3,-1,+1/2\rangle$ & $+$ & [\textbf{Rn}]$5f^{10}7s^2$ \\
99 & \textbf{Es}& $|5,3,0,+1/2\rangle$ & $+$ & [\textbf{Rn}]$5f^{11}7s^2$ \\
100 & \textbf{Fm}& $|5,3,1,+1/2\rangle$ & $+$ & [\textbf{Rn}]$5f^{12}7s^2$ \\
101 & \textbf{Md}& $|5,3,2,+1/2\rangle$ & $+$ & [\textbf{Rn}]$5f^{13}7s^2$ \\
102 & \textbf{No}& $|5,3,3,+1/2\rangle$ & $+$ & [\textbf{Rn}]$5f^{14}7s^2$ \\
103 & \textbf{Lr}& $|6,2,-2,-1/2\rangle$ & $+$ & [\textbf{Rn}]$5f^{14}6d^17s^2$ \\
104 & \textbf{Rf}& $|6,2,-1,-1/2\rangle$ & $+$ & [\textbf{Rn}]$5f^{14}6d^27s^2$ \\
105 & \textbf{Db}& $|6,2,0,-1/2\rangle$ & $+$ & [\textbf{Rn}]$5f^{14}6d^37s^2$ \\
106 & \textbf{Sg}& $|6,2,1,-1/2\rangle$ & $+$ & [\textbf{Rn}]$5f^{14}6d^47s^2$ \\
107 & \textbf{Bh}& $|6,2,2,-1/2\rangle$ & $+$ & [\textbf{Rn}]$5f^{14}6d^57s^2$ \\
108 & \textbf{Hs}& $|6,2,-2,+1/2\rangle$ & $+$ & [\textbf{Rn}]$5f^{14}6d^67s^2$ \\
109 & \textbf{Mt}& $|6,2,-1,+1/2\rangle$ & $+$ & [\textbf{Rn}]$5f^{14}6d^77s^2$ \\
110 & \textbf{Ds}& $|6,2,0,+1/2\rangle$ & $+$ & [\textbf{Rn}]$5f^{14}6d^87s^2$ \\
111 & \textbf{Rg}& $|6,2,1,+1/2\rangle$ & $+$ & [\textbf{Rn}]$5f^{14}6d^97s^2$ \\
112 & \textbf{Cn}& $|6,2,2,+1/2\rangle$ & $+$ & [\textbf{Rn}]$5f^{14}6d^{10}7s^2$ \\
113 & \textbf{Nh}& $|7,1,-1,-1/2\rangle$ & $+$ & [\textbf{Rn}]$5f^{14}6d^{10}7s^27p^1$ \\
114 & \textbf{Fl}& $|7,1,0,-1/2\rangle$ & $+$ & [\textbf{Rn}]$5f^{14}6d^{10}7s^27p^2$ \\
115 & \textbf{Mc}& $|7,1,1,-1/2\rangle$ & $+$ & [\textbf{Rn}]$5f^{14}6d^{10}7s^27p^3$ \\
116 & \textbf{Lv}& $|7,1,-1,+1/2\rangle$ & $+$ & [\textbf{Rn}]$5f^{14}6d^{10}7s^27p^4$ \\
117 & \textbf{Lv}& $|7,1,0,+1/2\rangle$ & $+$ & [\textbf{Rn}]$5f^{14}6d^{10}7s^27p^5$ \\
118 & \textbf{Og}& $|7,1,1,+1/2\rangle$ & $+$ & [\textbf{Rn}]$5f^{14}6d^{10}7s^27p^6$ \\
119 & \textbf{Uue}& $|8,0,0,-1/2\rangle$ & $+$ & [\textbf{Og}]$8s^1$ \\
120 & \textbf{Ubn}& $|8,0,0,+1/2\rangle$ & $+$ & [\textbf{Og}]$8s^2$ \\
\hline
\end{tabular}
}
\end{center}

\section{The Periodic Table and the Group $\SO(4,4)$}
The four quantum numbers $n$, $l$, $m$, $s$ included in the ket-vector (\ref{Ket1}) define four degrees of freedom. The first three quantum numbers $n$, $l$, $m$ have a clear geometric interpretation in Haenzel's Polygonfl\"{a}che \cite{Haenzel}  and the three-dimensional Finke system \cite{Finke}, see also \cite{Var2501,Var2502}. However, the fourth quantum number $s$ (spin) goes beyond the three-dimensional representation and requires the introduction of a fourth dimension for its adequate description\footnote{In the Haenzel and Finke systems, spin is artificially accounted for as points on the transversals. This clearly shows that the concept of spin does not fit into three-dimensional space, and all three-dimensional mechanical interpretations (like the Goudsmit-Uhlenbeck spinning top) are unable to describe spin by definition.}.

The theory of groups allows us to connect spin with the fourth dimension. Thus, according to the group-theoretic description of the periodic table \cite{Bar72,Fet,Var1801}, the first three quantum numbers $n$, $l$, and $m$ correspond to the eigenvalues $\nu$, $\lambda$, and $\mu_\lambda$ of the generators $\bsL_{56}$, $\bsL_{12}$ and $\bsL_{34}$ forming the Cartan subalgebra $\fK$ of the Lie algebra $\mathfrak{so}(4,2)$ of the conformal group $\SO(4,2)$. $\mathfrak{so}(4,2)$ is a Lie algebra of the third rank, therefore all root and weight diagrams of this algebra are three-dimensional systems. An adequate description of spin in the framework of a group-theoretic scheme requires a transition to a fourth-rank Lie algebra. Such an algebra is $\mathfrak{so}(4,4)$ -- the Lie algebra of the rotation group $\SO(4,4)$ of the eight-dimensional pseudo-Euclidean space $\R^{4,4}$\cite{Var2501,Var2502}.

A special pseudo-orthogonal group in eight dimensions, $\SO(4,4)$, corresponds to the rotation group of the eight-dimensional pseudo-Euclidean space $\R^{4,4}$, or, equivalently, the set of $8\times 8$ orthogonal matrices leaving a quadratic form
\[
Q(\boldsymbol{r})=x^2_1+x^2_2+x^2_3+x^2_4-x^2_5-x^2_6-x^2_7-x^2_8=\boldsymbol{r}^T\boldsymbol{r},
\]
where $\boldsymbol{r}=\ld x_1, x_2,x_3,x_4,x_5,x_6,x_7, x_8\rd^T$, invariant.

The structure of the corresponding Lie algebra $\mathfrak{so}(4,4)$ is determined by the commutation properties of its generators $\bsL_{\alpha\beta}$. $\bsL_{\alpha\beta}$ form the basis of the algebra $\mathfrak{so}(4,4)$. The number of independent generators is easy to find: out of 64 possible combinations of the indices $\alpha$ and $\beta$, eight combinations disappear by virtue of $\bsL_{\alpha\alpha}=0$, this reduces the number of generators to 56. Moreover, by virtue of $\bsL_{\alpha\beta}=-\bsL_{\beta\alpha}$ only 28 independent generators remain, the number of which can also be obtained using the formula $n(n-1)/2$, where $n=p+q$ is the dimension of the space $\R^{p,q}$. Thus,
\begin{equation}\label{LO2}
\bsL\Leftrightarrow\begin{bmatrix}
0 & \bsL_{12} & \bsL_{13} & \bsL_{14} & \bsL_{15} & \bsL_{16} & \bsL_{17} & \bsL_{18}\\
  & 0     & \bsL_{23}  & \bsL_{24} & \bsL_{25} & \bsL_{26} & \bsL_{27} & \bsL_{28}\\
  &       & 0      & \bsL_{34} & \bsL_{35} & \bsL_{36} & \bsL_{37} & \bsL_{38}\\
  &       &        & 0     & \bsL_{45}& \bsL_{46} & \bsL_{47} & \bsL_{48}\\
  &       &        &       & 0       & \bsL_{56} & \bsL_{57} & \bsL_{58}\\
  &       &        &       &         & 0         & \bsL_{67} & \bsL_{68}\\
  &       &        &       &         &           & 0         & \bsL_{78}\\
  &       &        &       &         &           &           & 0
\end{bmatrix}.
\end{equation}

The system of 28 generators $\bsL_{\alpha\beta}$ of the algebra $\mathfrak{so}(4,4)$ satisfies the following permutation relations:
\begin{equation}\label{Commut2}
\left[\bsL_{\alpha\beta},\bsL_{\gamma\delta}\right]=i\left(g_{\alpha\delta}\bsL_{\beta\gamma}+g_{\beta\gamma}\bsL_{\alpha\delta}
-g_{\alpha\gamma}\bsL_{\beta\delta}-g_{\beta\delta}\bsL_{\alpha\gamma}\right),
\end{equation}
where $\alpha,\beta,\gamma,\delta=1,\ldots,8$, $\alpha\neq\beta,\;\gamma\neq\delta$, while $g_{11}=g_{22}=g_{33}=g_{44}=1$, $g_{55}=g_{66}=g_{77}=g_{88}=-1$. Thus, we have a 28-dimensional Lie algebra $\mathfrak{so}(4,4)$.

Let's find the maximal subset of commuting generators of the algebra $\mathfrak{so}(4,4)$. As is known, two generators commute if they do not have common indexes. It is easy to see that among the generators of the algebra $\mathfrak{so}(4,4)$, four generators satisfy this condition:
\begin{equation}\label{Cartan}
\bsL_{12},\;\bsL_{34},\;\bsL_{56},\;\bsL_{78}.
\end{equation}
The four generators $\lf\bsL_{12},\bsL_{34},\bsL_{56},\bsL_{78}\rf$ form the basis of \textit{maximal abelian subalgebra} $\fK\subset\mathfrak{so}(4,4)$ (\textit{Cartan subalgebra}). The generators (\ref{Cartan}) are called \textit{Cartan generators}. The dimension of the subalgebra $\fK$ defines the \textit{rank} of the Lie algebra, hence $\mathfrak{so}(4,4)$ is a Lie algebra of the fourth rank. Thus, all root and weight diagrams for $\mathfrak{so}(4,2)$ will be four-dimensional\footnote{As a consequence, this excludes explicit visualization of these diagrams.}.

\begin{figure}[ht] %
\centering
\includegraphics[width=15cm]{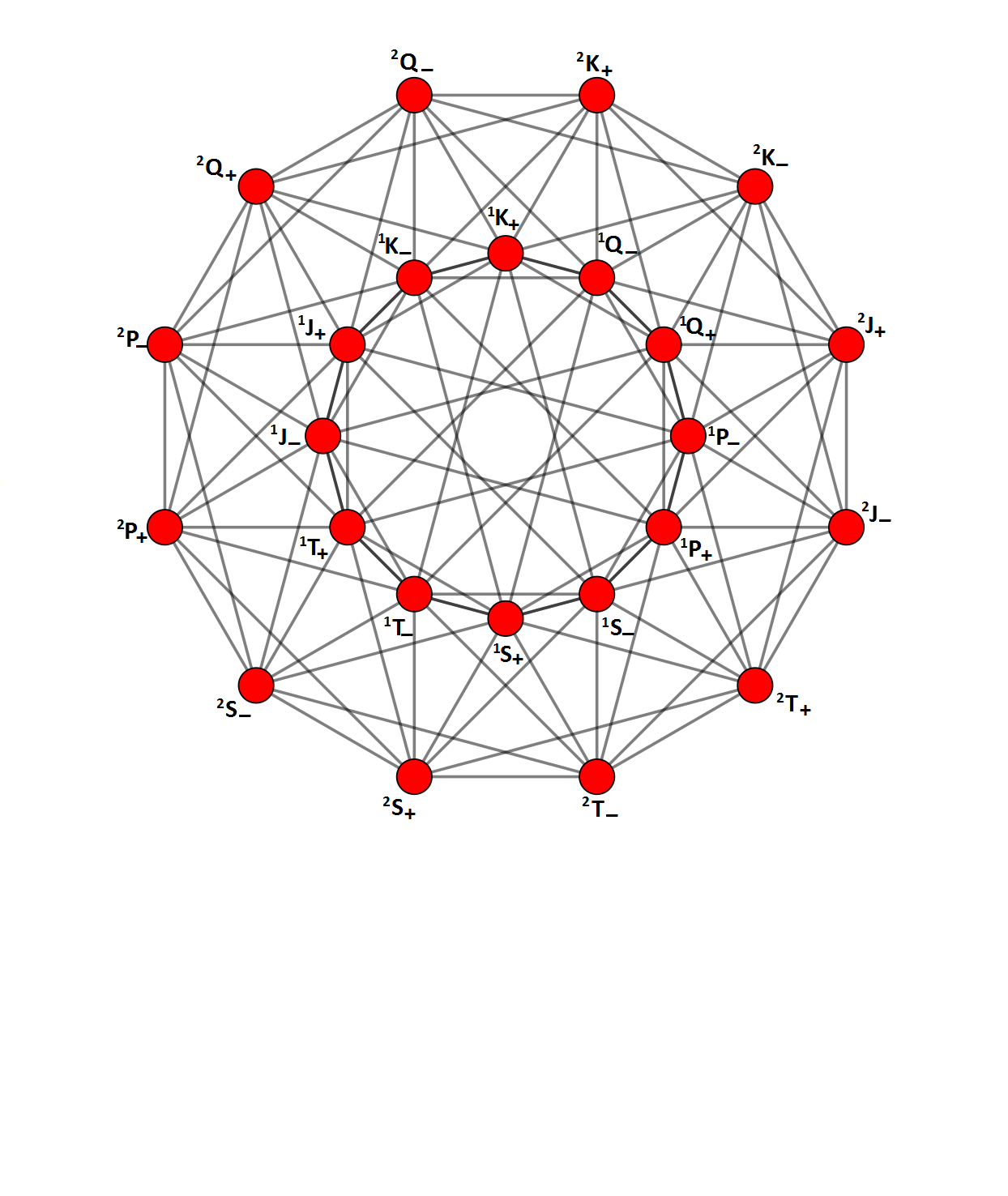}\\
\vspace{-5cm}
\caption{Root diagram of the Lie algebra $\mathfrak{so}(4,4)$. Weyl generators form the vertices of a 24-cell in four-dimensional space. The figure shows an orthogonal projection of an octaplex onto a two-dimensional plane.}
\label{pic_Octaplex}%
\end{figure}

The \textit{Cartan-Weyl basis} of the algebra $\mathfrak{so}(4,4)$ contains 28 generators,
\begin{multline}
\left\{\bsL_{12},\bsL_{34},\bsL_{56},\bsL_{78},{}^1\bsK_+,{}^1\bsK_-,{}^1\bsJ_+,{}^1\bsJ_-,{}^1\bsT_+,{}^1\bsT_-,
{}^1\bsS_+,{}^1\bsS_-,{}^1\bsP_+,{}^1\bsP_-,{}^1\bsQ_+,{}^1\bsQ_-,\right.\\
\left.{}^2\bsK_+,{}^2\bsK_-,{}^2\bsJ_+,{}^2\bsJ_-,{}^2\bsT_+,{}^2\bsT_-,
{}^2\bsS_+,{}^2\bsS_-,{}^2\bsP_+,{}^2\bsP_-,{}^2\bsQ_+,{}^2\bsQ_-\right\},\label{CWbasis}
\end{multline}
including 4 Cartan generators and 24 Weyl generators. Thus, the root diagram of the algebra $\mathfrak{so}(4,4)$ is defined in a four-dimensional weight space whose coordinate axes are the Cartan generator axes, and the 24 Weyl generator axes form a regular polyhedron in four-dimensional space. Figure \ref{pic_Octaplex} shows the root diagram of $\mathfrak{so}(4,4)$ as a 24-cell. A 24-cell or \textit{octaplex} (other names: polyoctahedron, icositetrachoron) is the convex regular self-dual polyhedron (polytope) in four-dimensional space. The octaplex contains 24 vertices, 96 edges, 96 triangular faces, and 24 octahedral cells with six meeting at each vertex, and three at each edge. The symmetry group $F_4$ (Coxeter group \cite{Cox}) of this polyhedron has the order 1152. An octaplex is the only self-dual regular polytope of dimension greater than 2 that is not a simplex. This is the reason for the uniqueness of the octaplex: unlike the other five regular four-dimensional polyhedra (pentachoron, tesseract, hexadecachoron, hypericosahedron, hyperdodecahedron), it has no analogue among the five Platonic solids (regular three-dimensional polyhedra). The three-dimensional orthographic projection of the octaplex is a \textit{cuboctahedron} (one of the 13 Archimedean solids). It should be noted that the four-dimensional space contains the largest number of different types of regular polyhedra, equal to 6. For all other $n$-dimensional spaces, for $n>4$, this number is 3.

\begin{figure}[ht] %
\centering
\includegraphics[width=16cm]{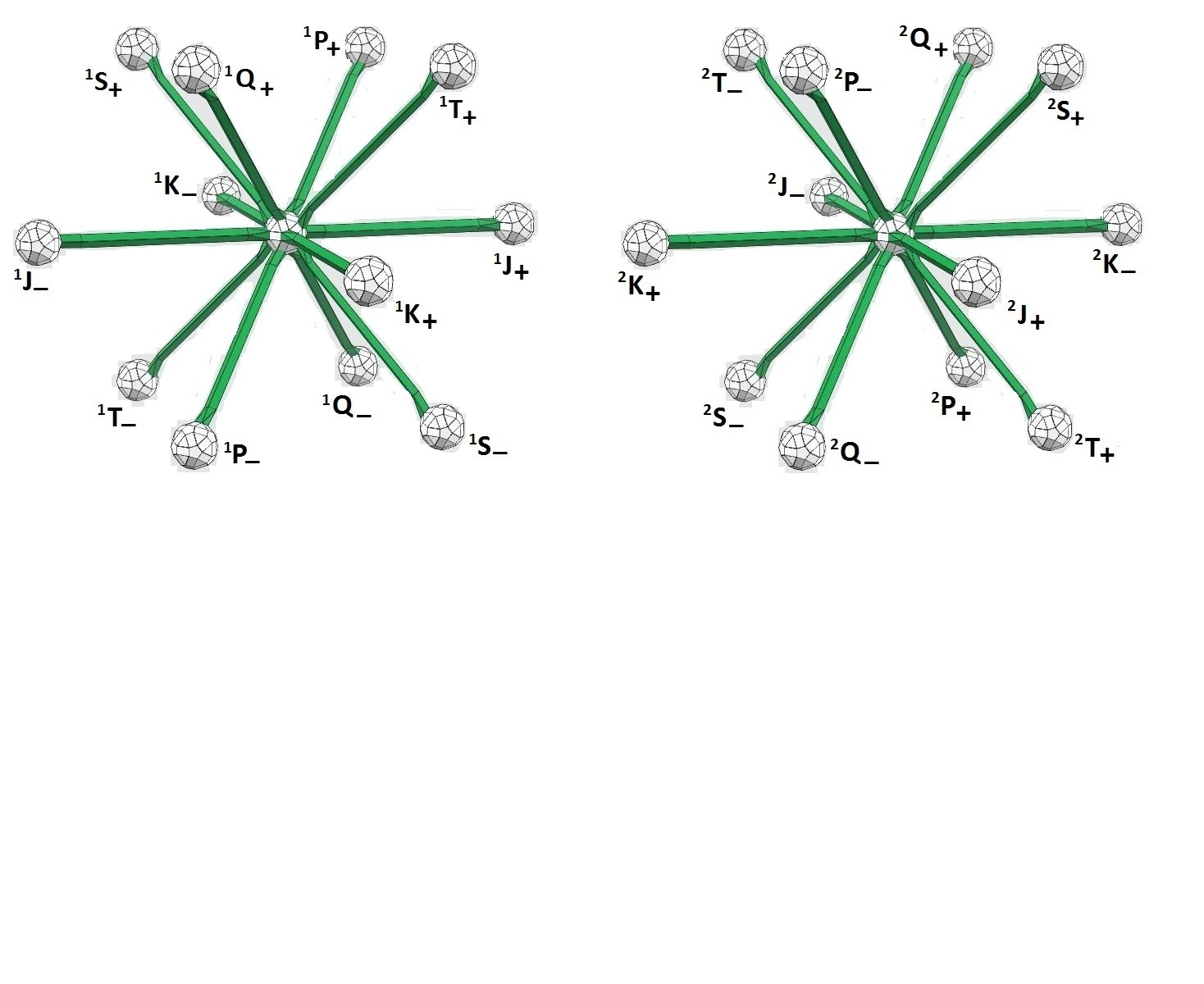}\\
\vspace{-6cm}
\caption{Root diagrams (cuboctahedra) of split bases (\ref{IB1}) and (\ref{IB2}) of the Lie algebra $\mathfrak{so}(4,4)$.}
\label{pic2}%
\end{figure}

The fourth generator $\bsL_{78}$, understood as a spin generator\footnote{The eigenvalue $\sigma$ of the generator $\bsL_{78}$ corresponds to the fourth quantum number $s$.}, commutes with all 15 generators of the subalgebra $\mathfrak{so}(4,2)$. As a consequence, the Cartan-Weyl basis (\ref{CWbasis}) for the algebra $\mathfrak{so}(4,4)$ splits into two structurally identical bases
\begin{equation}\label{IB1}
\left\{{}^1\bsK_3,{}^1\bsJ_3,{}^1\bsT_0,{}^1\bsS_0,{}^1\bsP_0,{}^1\bsQ_0,{}^1\bsK_+,{}^1\bsK_-,{}^1\bsJ_+,
{}^1\bsJ_-,
{}^1\bsT_+,{}^1\bsT_-,{}^1\bsS_+,{}^1\bsS_-,{}^1\bsP_+,{}^1\bsP_-,{}^1\bsQ_+,{}^1\bsQ_-\right\},
\end{equation}
\begin{equation}\label{IB2}
\left\{{}^2\bsK_3,{}^2\bsJ_3,{}^2\bsT_0,{}^2\bsS_0,{}^2\bsP_0,{}^2\bsQ_0,{}^2\bsK_+,{}^2\bsK_-,{}^2\bsJ_+,
{}^2\bsJ_-,
{}^2\bsT_+,{}^2\bsT_-,{}^2\bsS_+,{}^2\bsS_-,{}^2\bsP_+,{}^2\bsP_-,{}^2\bsQ_+,{}^2\bsQ_-\right\},
\end{equation}
each of which is isomorphic to the Yao basis \cite{Yao1} for the group algebra of the twofold covering $\spin_+(4,2)\simeq\SU(2,2)$. The bases (\ref{IB1}) and (\ref{IB2}) define two root systems
\begin{equation}\label{RS1}
\ar\left.
\begin{array}{ccc}
\boldsymbol{\alpha}({}^1\bsK_+)=(+1,+1,0),&\;\boldsymbol{\alpha}({}^1\bsT_+)=(+1,0,+1),&
\quad\boldsymbol{\alpha}({}^1\bsP_+)=(0,+1,+1),\\
\boldsymbol{\alpha}({}^1\bsK_-)=(-1,-1,0),&\;\boldsymbol{\alpha}({}^1\bsT_-)=(-1,0,-1),&
\quad\boldsymbol{\alpha}({}^1\bsP_-)=(0,-1,-1),\\
\boldsymbol{\alpha}({}^1\bsJ_+)=(-1,+1,0),&\;\boldsymbol{\alpha}({}^1\bsS_+)=(-1,0,+1),&
\quad\boldsymbol{\alpha}({}^1\bsQ_+)=(0,-1,+1),\\
\boldsymbol{\alpha}({}^1\bsJ_-)=(+1,-1,0),&\;\boldsymbol{\alpha}({}^1\bsS_-)=(+1,0,-1),&
\quad\boldsymbol{\alpha}({}^1\bsQ_-)=(0,+1,-1),\\
\end{array}\right\}
\end{equation}
\begin{equation}\label{RS2}
\ar\left.
\begin{array}{ccc}
\boldsymbol{\alpha}({}^2\bsK_+)=(+1,-1,0),&\;\boldsymbol{\alpha}({}^2\bsT_+)=(+1,0,-1),&
\quad\boldsymbol{\alpha}({}^2\bsP_+)=(0,+1,-1),\\
\boldsymbol{\alpha}({}^2\bsK_-)=(-1,+1,0),&\;\boldsymbol{\alpha}({}^2\bsT_-)=(-1,0,+1),&
\quad\boldsymbol{\alpha}({}^2\bsP_-)=(0,-1,+1),\\
\boldsymbol{\alpha}({}^2\bsJ_+)=(+1,+1,0),&\;\boldsymbol{\alpha}({}^2\bsS_+)=(+1,0,+1),&
\quad\boldsymbol{\alpha}({}^2\bsQ_+)=(0,+1,+1),\\
\boldsymbol{\alpha}({}^2\bsJ_-)=(-1,-1,0),&\;\boldsymbol{\alpha}({}^2\bsS_-)=(-1,0,-1),&
\quad\boldsymbol{\alpha}({}^2\bsQ_-)=(0,-1,-1),\\
\end{array}\right\}
\end{equation}
\[
\boldsymbol{\alpha}(\sL_3)=(0,0,0),\quad\boldsymbol{\alpha}(\sA_3)=(0,0,0),
\quad\boldsymbol{\alpha}(\Delta_3)=(0,0,0).
\]
Graphically, the root systems (\ref{RS1}) and (\ref{RS2}) can be represented by two cuboctahedra shown in Figure \ref{pic2}.

\begin{figure}[p] %
\centering
\vspace{-0.5cm}
\includegraphics[width=\textwidth]{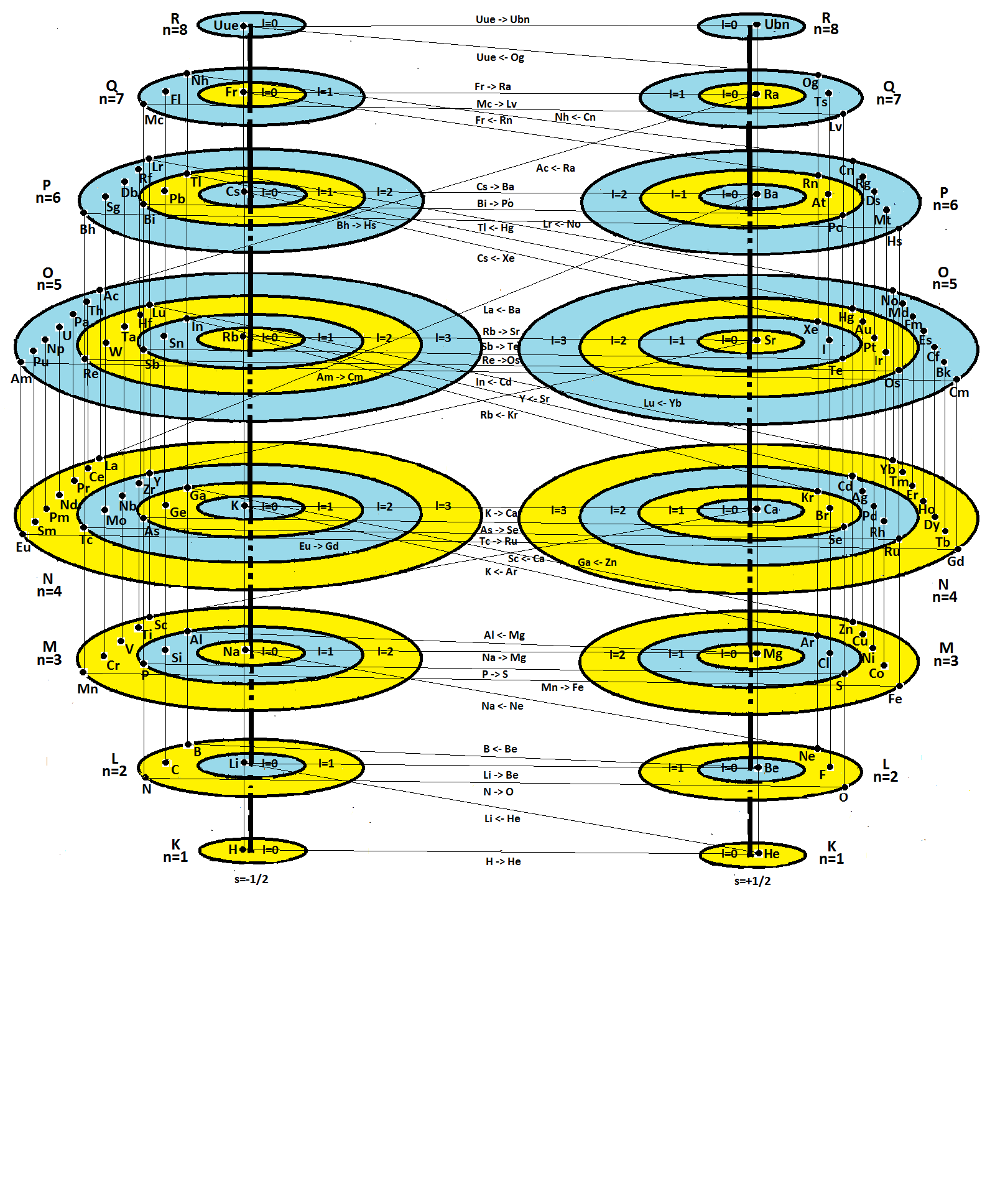}\\
\vspace{-5cm}
\caption{The periodic system of chemical elements in the split basis of the Lie algebra $\mathfrak{so}(4,4)$.}
\label{pic3}%
\end{figure}

The above reduction of the root diagram of the algebra $\mathfrak{so}(4,4)$ leads to a similar reduction for the weight diagram. Figure \,\ref{pic3} shows two three-dimensional projections ($\SO(4,2)$-towers) of the weight diagram of the Lie algebra $\mathfrak{so}(4,4)$ of the rotation group $\SO(4,4)$ of an eight-dimensional pseudo-Euclidean space $\R^{4,4}$, corresponding to the root systems (\ref{RS1}) and (\ref{RS2}). The vertical axes of each $\SO(4,2)$-tower are formed by the eigenvalues of the Cartan generator $\boldsymbol{\Delta}_3=\bsL_{56}$, which adds a radial ladder operator to the variety. Each given floor of a $\SO(4,2)$ tower (Haenzel circle) is characterized by the main quantum number $n$. The horizontal bands (on the floors) correspond to various $l$-subshells (Haenzel rings), and the points are individual $m$-components (finite-dimensional representations of the group $\SO(4,2)$) defining the elements of the periodic table according to the Madelung rule. Rings containing elements with an odd sum $n+l$ are indicated in yellow, respectively, rings with an even sum $n+l$ are indicated in blue. Homologous elements are connected by vertical lines (Bailey-Thomsen-Bohr lines). Transitions between different quantum levels are indicated by directional lines (Haenzel lines). On the radial axes of the $\SO(4,2)$-towers are metals with quantum numbers ($n,l=0,s=\pm 1/2$), $n=1,2,\ldots$. On the radial axis of the $\SO(4,2)$-tower with the quantum number $s=-1/2$ (the left tower) are the alkali metals of group I (\textbf{Li}, \textbf{Na}, \textbf{K}, \textbf{Rb}, \textbf{Cs}, \textbf{Fr}, $\ldots$), respectively, on the right tower ($s=+1/2$), the radial axis is inhabited by group II alkaline earth metals (\textbf{Be}, \textbf{Mg}, \textbf{Ca}, \textbf{Sr}, \textbf{Ba}, \textbf{Ra}, $\ldots$). As we move away from the radial axes to the periphery of the Haenzel circles, the \textit{metallicity} of the elements decreases and \textit{nonmetallicity} increases, which corresponds to a movement from left to right along the period in the standard periodic table (metals $\rightarrow$ amphoteric elements $\rightarrow$ inert gases).

The first period of the periodic table, which includes hydrogen \textbf{H} and helium \textbf{He}, corresponds to the two starting points on the diagram (the first floor $n=1$ of $\SO(4,2)$-towers), from which the sequences of elements with $s=-1/2$ and $s=+1/2$ begin, respectively. The first period is the only period that does not contain a double. The second element of the 1st period, helium \textbf{He}, is inert and completes the period. In this regard, the question arises: where in the 1st period is that natural change of properties from metallicity to nonmetallicity (from pole to pole)? The thing is that the first element of the 1st period, hydrogen \textbf{H}, is an amphoteric element to some extent. By some properties (for example, it is monovalent and forms a singly charged positive ion $\mathbf{H}^+$), hydrogen is close to the alkali metals of group I. Therefore, in some editions of the periodic table \textbf{H} is placed in this group. According to other properties (for example, starting from the ability to form a negative singly charged ion $\mathbf{H}^-$ and ending with the physical properties of hydrogen gas $\mathbf{H}_2$), it is closer to the halogen elements of group VII. We can say that hydrogen \textbf{H} combines different ``pole position'' properties\footnote{For this reason, the symbol of hydrogen \textbf{H} appears twice in the periodic table: in both groups I and VII. Usually, in one of these groups, his symbol is enclosed in parentheses, wanting to show his lesser right to be in this group. The debate about the true place of hydrogen in the periodic table has a long history. From the standpoint of the present group-theoretic approach, hydrogen \textbf{H} belongs to the alkali metals of group I.}. Haenzel line \textbf{He} $\rightarrow$ \textbf{Li} shows the transition to the next ($n=2$) quantum level.

The second period (floor $n=2$) contains lithium \textbf{Li} and beryllium \textbf{Be} connected by the line \textbf{Li} $\rightarrow$ \textbf{Be}. Further down the line \textbf{Be} $\rightarrow$ \textbf{B} there is a transition to the amphoteric elements \textbf{B}, \textbf{C}, \textbf{N}, the Haenzel ring ($n=2,l=1,s=-1/2$), after which the transition follows \textbf{N}$\rightarrow$ \textbf{O} and filling the ring ($n=2,l=1,s=+1/2$) to an inert gas \textbf{Ne}, thereby completing the second period and the complete occupation of the 2nd floor of the $\SO(4,2)$-towers.

The third period begins with the transition \textbf{Ne} $\rightarrow$ \textbf{Na} (transition to the 3rd floor $n=3$) from neon \textbf{Ne} to alkali metal \textbf{Na}. This is followed by the transition \textbf{Na} $\rightarrow$ \textbf{Mg} and the rings ($n=3,l=1,s=-1/2$) and ($n=3,l=1,s=+1/2$) are filled with amphoteric elements \textbf{Al}, \textbf{Si}, $\ldots$ before argon \textbf{Ar} (the third period is a double of the second period). The third period has been completed, but the 3rd floor of the $\SO(4,2)$-towers is still not fully occupied.

The fourth period begins with the transition \textbf{Ar} $\rightarrow$ \textbf{K} (transition to the 4th floor $n=4$ of the left tower $s=-1/2$) from the inert gas \textbf{Ar} to the alkali metal \textbf{K} and then there is the transition \textbf{K}$\rightarrow$ \textbf{Ca}, after which the return to the 3rd floor takes place (\textbf{Ca} $\rightarrow$ \textbf{Sc}) and filling of the ring ($n=3,l=2,s=-1/2$) with transition metals of side subgroups from scandium \textbf{Sc} to manganese \textbf{Mn}. Then, after the transition \textbf{Mn} $\rightarrow$ \textbf{Fe} fills the ring ($n=3,l=2,s=+1/2$) of the right tower with transition metals of side subgroups, starting with iron \textbf{Fe} and ending with zinc \textbf{Zn}. This is followed by the transition \textbf{Zn} $\rightarrow$ \textbf{Ga} and the occupation of the 4th floor with amphoteric elements (semimetals) \textbf{Ga}, \textbf{Ge}, \textbf{As} rings ($n=4,l=1,s=-1/2$) continues. After switching to \textbf{As} $\rightarrow$ \textbf{Se} the fourth period ends with the filling of the ring ($n=4,l=1,s=+1/2$) with amphoteric elements \textbf{Se}, \textbf{Br}, \textbf{Kr}. The fourth period has been completed, but the 4th floor of the $\SO(4,2)$-towers remains not fully occupied.

The fifth period is a double of the fourth and, like the 4th, contains 18 elements. Accordingly, the structure of its filling is similar to the fourth period. \textbf{Kr} $\rightarrow$ \textbf{Rb} defines the transition to the 5th floor of the left $\SO(4,2)$-tower from the inert gas \textbf{Kr} to the alkali metal \textbf{Rb}. This is followed by the return \textbf{Rb} $\rightarrow$ \textbf{Sr} to the right tower, followed by the transition \textbf{Sr} $\rightarrow$ \textbf{Y} to the 4th floor of the left tower and filling the ring ($n=4,l=2,s=-1/2$) with transition metals \textbf{Y}, \textbf{Zr}, $\ldots$, \textbf{Tc}. This is followed by the transition \textbf{Tc} $\rightarrow$ \textbf{Ru} and the filling of the ring ($n=4,l=2,s=+1/2$), starting from the transition metal \textbf{Ru} to cadmium \textbf{Cd}. Transition \textbf{Cd} $\rightarrow$ \textbf{In} sets the return to the 5th floor of the left tower, after which the ring ($n=5,l=1,s=-1/2$) is filled with post-transition metals (semimetals) \textbf{In}, \textbf{Sn}, \textbf{Sb}. After switching to \textbf{Sb} $\rightarrow$ \textbf{Te} the fifth period ends with the filling of the ring ($n=5,l=1,s+1/2$) from the metalloid \textbf{Te} to the inert gas \textbf{Xe}. The fifth period has been completed, but the 4th and 5th floors of the $\SO(4,2)$-towers remain not yet fully occupied.

The sixth period begins with the transition \textbf{Xe} $\rightarrow$ \textbf{Cs} (transition to the 6th floor of the left tower) from xenon \textbf{Xe} to alkali metal \textbf{Cs}. Then after moving to the 6th floor of the right tower \textbf{Cs} $\rightarrow$ \textbf{Ba} going downhill \textbf{Ba} $\rightarrow$ \textbf{La} to the 4th floor of the left tower and filling the ring ($n=4,l=3,s=-1/2$) with lanthanides (rare earth metals) \textbf{La}, \textbf{Ce}, $\ldots$, \textbf{Eu}. After switching to \textbf{Eu} $\rightarrow$ \textbf{Gd} on the 4th floor of the right tower, the lanthanide family continues to fill \textbf{Gd}, \textbf{Tb}, $\ldots$, \textbf{Yb}. The 4th floor of the $\SO(4,2)$-towers is fully occupied. The lanthanide family fills the outer rings of the 4th floor. Lifting \textbf{Yb} $\rightarrow$ \textbf{Lu} on the 5th floor of the left tower, the settlement of the ring ($n=5,l=2,s=-1/2$) with transition metals continues, starting with lutetium\footnote{There is a duality in the classification of the element \textbf{Lu}. On the one hand, lutetium \textbf{Lu} is considered to be the last element in the lanthanide series, i.e. it belongs to rare earth metals. On the other hand, it can also be classified as the first element of the transition metals of the 6th period. According to the proposed interpretation and as follows from the diagram in Figure \ref{pic3}, lutetium \textbf{Lu} is a transition metal of the 6th period. The lanthanide family contains 14 elements \textbf{La}, \textbf{Ce}, $\ldots$, \textbf{Yb} and is located on the outer rings ($n=4,l=3,s=-1/2$) and ($n=4,l=3,s+1/2$) of the 4th floor of $\SO(4,2)$-towers.} \textbf{Lu} to rhenium \textbf{Re}. Switching to osmium \textbf{Os}, \textbf{Re} $\rightarrow$ \textbf{Os}, begins to populate the ring ($n=5,l=2,s=+1/2$) with the remaining transition metals of the 6th period: \textbf{Os}, \textbf{Ir}, $\ldots$, \textbf{Hg}. Lifting \textbf{Hg} $\rightarrow$ \textbf{Tl} continues filling the 6th floor of the left tower with heavy post-transition metals \textbf{Tl}, \textbf{Pb}, \textbf{Bi} in the ring ($n=6,l=1,s=-1/2$). After switching to \textbf{Bi} $\rightarrow$ \textbf{Po} the sixth period ends with the filling of the ring ($n=6,l=1,s=+1/2$) from the semimetal \textbf{Po} to radon (inert gas) \textbf{Rn}. The sixth period has been completed, but the 5th and 6th floors of the $\SO(4,2)$-towers remain not fully occupied.

The seventh period (the last period of the periodic table) is a double of the sixth and, like the sixth, contains 32 elements. The period begins with transition \textbf{Rn} $\rightarrow$ \textbf{Fr} from inert gas \textbf{Rn} to alkali metal \textbf{Fr} (francium) on the 7th floor of the left tower. This is followed by the transition \textbf{Fr} $\rightarrow$ \textbf{Ra} to the right tower and descent \textbf{Ra} $\rightarrow$ \textbf{Ac} on the 5th floor of the left tower to the family of actinoids \textbf{Ac}, \textbf{Th}, $\ldots$, \textbf{Am}, inhabiting the ring ($n=5,l=3,s=-1/2$). Next, go to \textbf{Am} $\rightarrow$ \textbf{Cm} on the 5th floor of the right tower, the actinoid family continues to be filled with elements \textbf{Cm}, \textbf{Bk}, $\ldots$, \textbf{No} in the ring ($n=5,l=3,s=+1/2$). 5th floor of the $\SO(4,2)$-towers are inhabited\footnote{However, unlike the previous (sixth) period, in which the lanthanide family completely inhabited the 4th floor, in the case of the seventh period, the 5th floor remains \textit{not fully populated}, namely, the rings  ($n=5,l=4,s=-1/2$) and ($n=5,l=4,s=+1/2$) are empty. These rings are filled with hypothetical elements of the eighth period of the Seaborg table (for more details, see section 5).}. The actinoid family fills the outer rings of the 5th floor and contains 14 elements. The elements of this family are homologues of the elements of the lanthanide family located below on the 4th floor. The homology of the elements of these families is marked by vertical lines (Bailey-Thomsen-Bohr lines). Entrance \textbf{No} $\rightarrow$ \textbf{Lr} on the 6th floor of the left tower, the settlement of the ring ($n=6,l=2,s=-1/2$) with transition metals continues, starting with lawrencium\footnote{As in the case of lutetium \textbf{Lu}, there is a duality in the classification of the element \textbf{Lr}. On the one hand, it is believed that lawrencium \textbf{Lr} is the last element in the series of actinides, and on the other hand, it can also be classified as the first element of transition metals of the 7th period. According to the diagram in Figure \ref{pic3}, lawrencium \textbf{Lr} is the first transition metal of the 7th period, and it is also a homologue of lutetium \textbf{Lu} (the first transition metal of the 6th period), which is indicated on the diagram by the corresponding vertical line. Thus, the actinoid family \textbf{Ac}, \textbf{Th}, $\ldots$, \textbf{No} contains 14 elements and is located on the outer rings ($n=5,l=3,s=-1/2$) and ($n=5,l=3,s=+1/2$) of the 5th floor of $\SO(4,2)$ towers, being, at the same time, a homologue of the lanthanide family.} \textbf{Lr} and ending with borium \textbf{Bh}. Transition to hassium \textbf{Hs}, \textbf{Bh} $\rightarrow$ \textbf{Hs}, begins to populate the ring ($n=6,l=2,s=+1/2$) with the remaining transition metals of the 7th period: \textbf{Hs}, \textbf{Mt}, $\ldots$, \textbf{Cn}. Lifting \textbf{Cn} $\rightarrow$ \textbf{Nh} continues filling the 7th floor of the left tower with heavy post-transition metals of the 7th period in the ring ($n=7,l=1,s=-1/2$): nihonium \textbf{Nh}, flerovium \textbf{Fl}, moscovium \textbf{Mc}. After switching to \textbf{Mc} $\rightarrow$ \textbf{Lv} the seventh period ends with the filling of the ring ($n=7,l=1,s=+1/2$) from the post-transition metal \textbf{Lv} (livermorium) to the oganesson (inert gas) \textbf{Og}. The seventh period is over. However, the 5th, 6th, and 7th floors of the $\SO(4,2)$-towers remain not fully occupied.

Transition \textbf{Og} $\rightarrow$ \textbf{Uue} from an inert gas \textbf{Og} to a hypothetical element\footnote{According to the logic of the diagram in Figure \ref{pic3}, \textbf{Uue} -- alkali metal.} \textbf{Uue} begins the eighth period of the periodic table of elements, which already goes beyond the limits of the usual seven-period table. The Seaborg table (see section 5) begins with the eighth period, the elements of which fill the empty rings of the 5th, 6th and 7th floors of the $\SO(4,2)$ towers. And then the towers go to infinity, and on both sides of the plane ($\sL_3,\sA_3$), thereby forming pyramids of matter and antimatter.

\section{The structure of the levels}
The geometry of the periodic table is determined by the ordinal structure of quantum numbers.
\begin{figure}[p] %
\centering
\includegraphics[width=16cm]{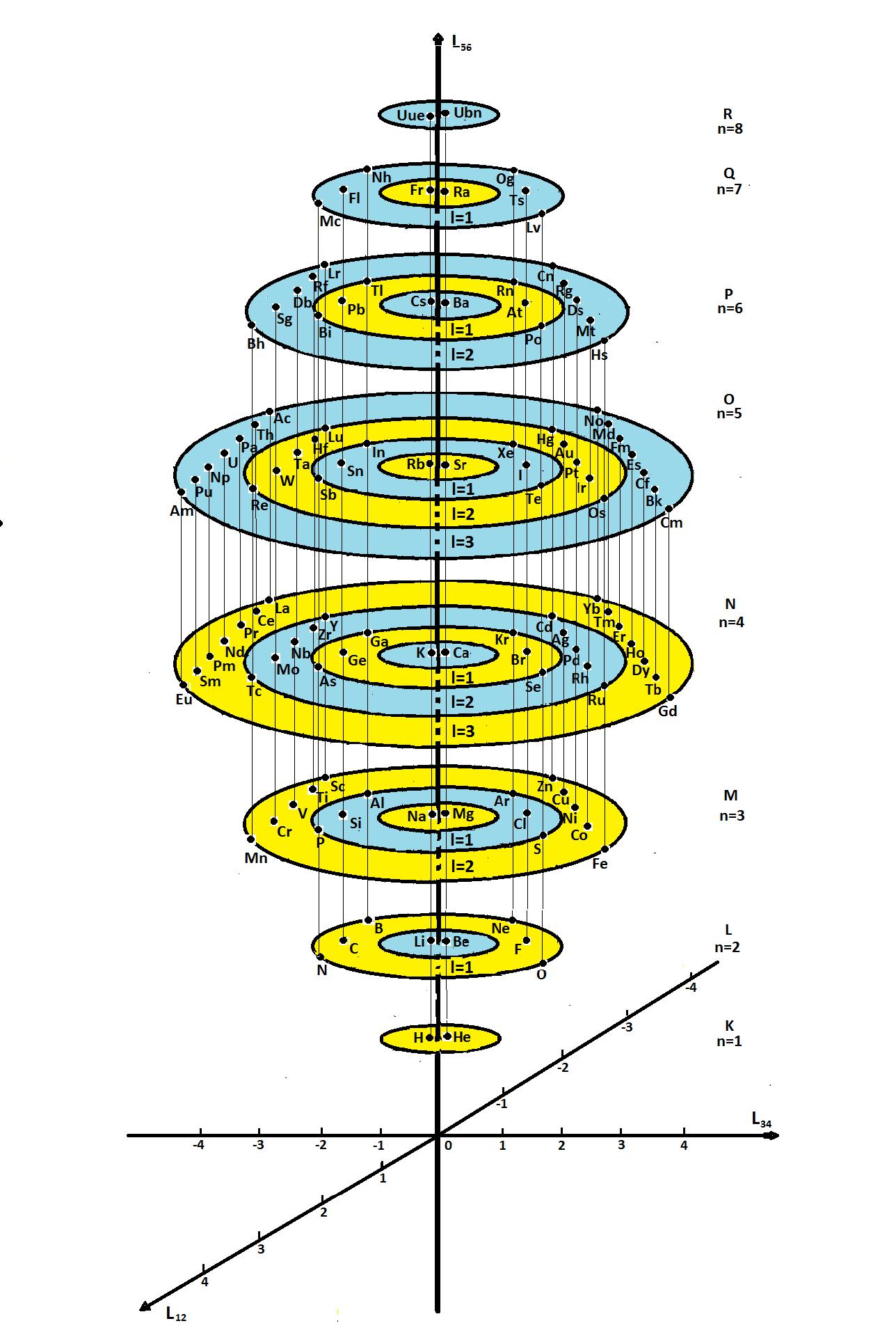}\\
\caption{Periodic system of chemical elements in the form of a combined weight diagram of split bases (\ref{IB1}) and (\ref{IB2}) of the Lie algebra $\mathfrak{so}(4,4)$.}
\label{pic5}%
\end{figure}
Figure \ref{pic5} shows two combined three-dimensional projections ($\SO(4,2)$-towers) of the weight diagram of the Lie algebra $\mathfrak{so}(4,4)$ of the rotation group $\SO(4,4)$ corresponding to the bases (\ref{IB1}) and (\ref{IB2}).

\begin{figure}[ht] %
\centering
\includegraphics[width=\textwidth]{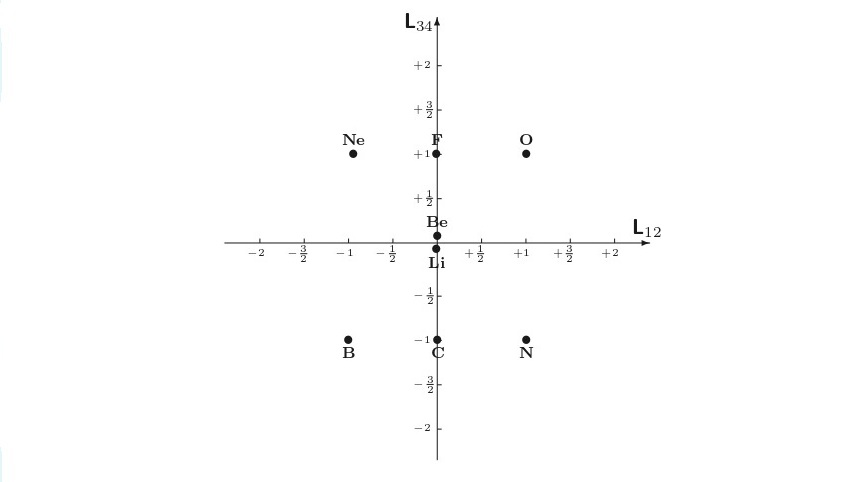}\\
\caption{The level $n=2$ of the weight diagram of the algebra $\mathfrak{so}(4,4)$ contains the $(1,1)$-diagram of the subalgebra $\mathfrak{so}(4)$, which completely includes the second period.}
\label{multiplet_n2}%
\end{figure}

\begin{figure}[ht] %
\centering
\includegraphics[width=16cm]{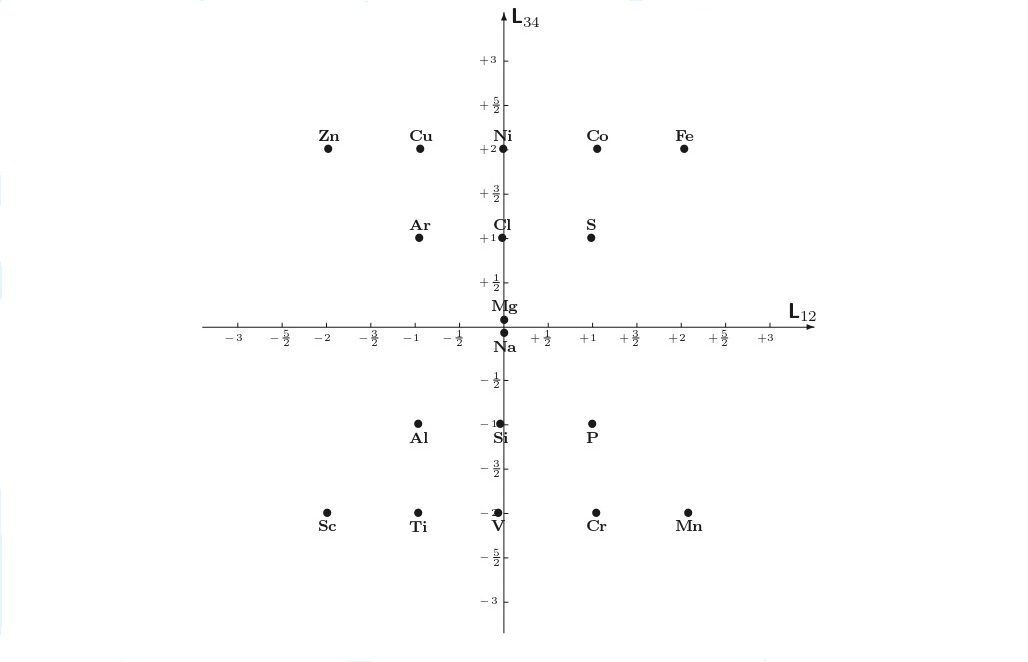}\\
\caption{The level $n=3$: the $(2,2)$-diagram of the subalgebra $\mathfrak{so}(4)$ contains the entire 3rd period and transition metals of the 4th period \textbf{Sc}, $\ldots$, \textbf{Mn}, \textbf{Fe}, $\ldots$, \textbf{Zn}.}
\label{multiplet_n3}%
\end{figure}

\begin{figure}[ht] %
\centering
\includegraphics[width=16cm]{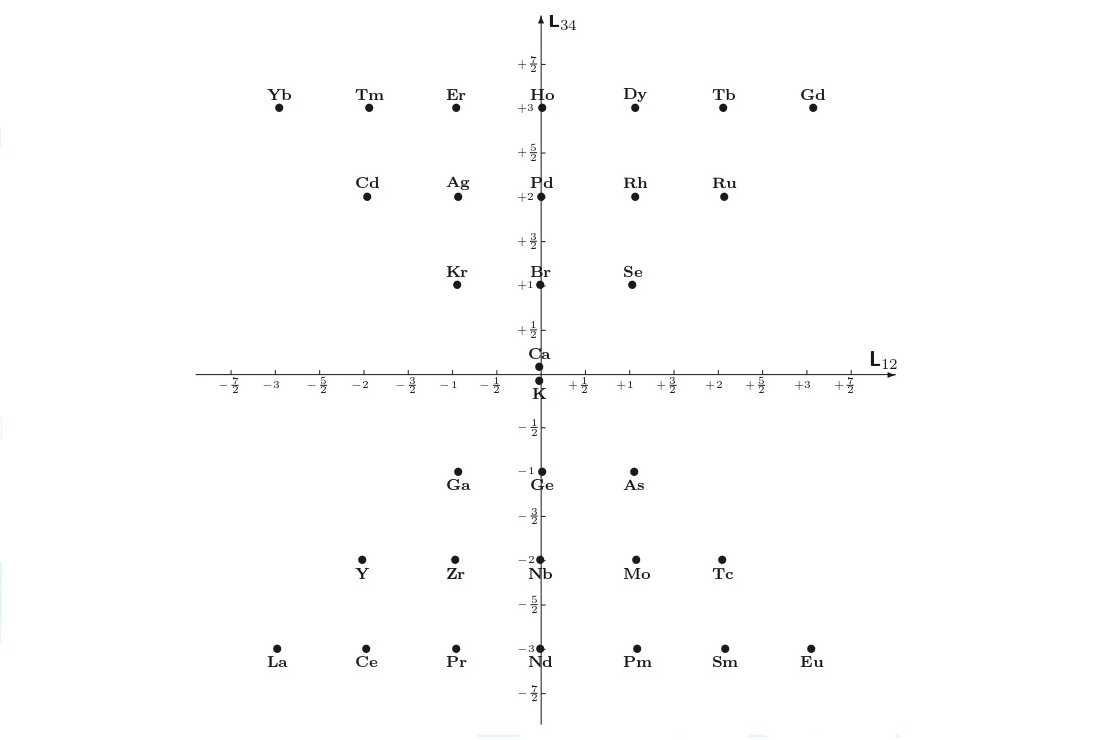}\\
\caption{The level $n=4$: the $(3,3)$-diagram of the subalgebra $\mathfrak{so}(4)$ contains the completion of the 4th period, transition metals of the 5th period \textbf{Y}, $\ldots$, \textbf{Tc}, \textbf{Ru}, $\ldots$, \textbf{Cd}, as well as the lanthanide family \textbf{La}, \textbf{Ce}, $\ldots$, \textbf{Yb}.}
\label{multiplet_n4}%
\end{figure}

\begin{figure}[ht] %
\centering
\includegraphics[width=15cm]{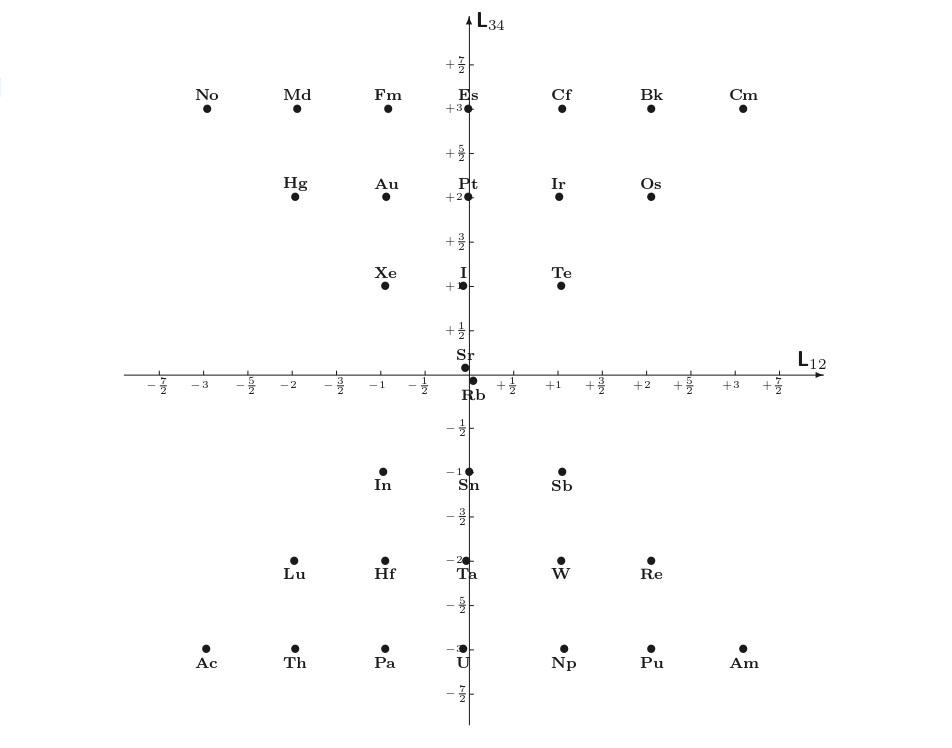}\\
\caption{The level $n=5$: the $(3,3)$-diagram of the subalgebra $\mathfrak{so}(4)$ contains the completion of the 5th period, the transition metals of the 6th period \textbf{Lu}, $\ldots$, \textbf{Re}, \textbf{Os}, $\ldots$, \textbf{Hg}, as well as the actinoid family \textbf{Ac}, \textbf{Th}, $\ldots$, \textbf{No}.}
\label{multiplet_n5}%
\end{figure}

\begin{figure}[ht] %
\centering
\includegraphics[width=16cm]{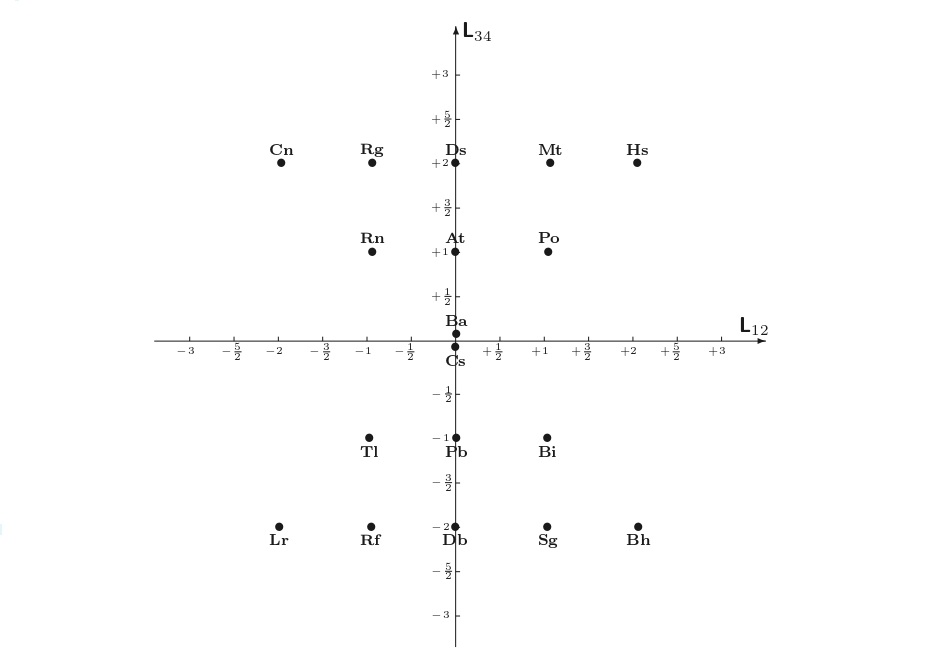}\\
\caption{The level $n=6$: the $(2,2)$-diagram of the subalgebra $\mathfrak{so}(4)$ contains the completion of the 6th period and the transition metals of the 7th period \textbf{Lr}, $\ldots$, \textbf{Bh}, \textbf{Hs}, $\ldots$, \textbf{Cn}.}
\label{multiplet_n6}%
\end{figure}

\begin{figure}[ht] %
\centering
\includegraphics[width=16cm]{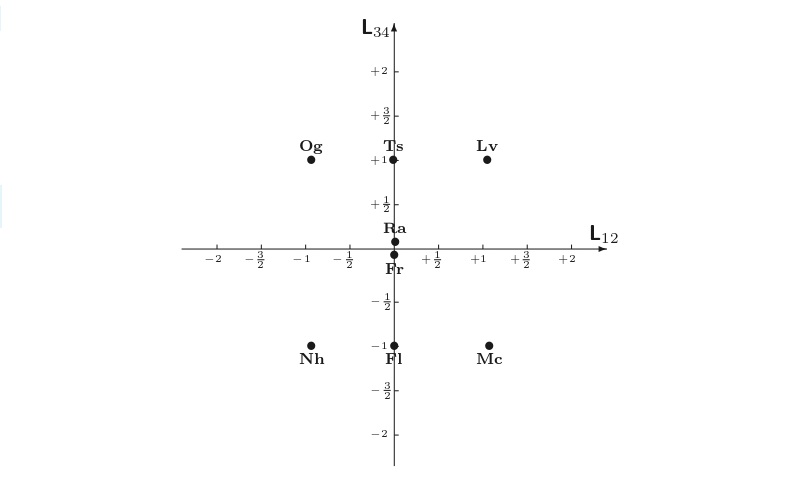}\\
\caption{The level $n=7$: the $(1,1)$-diagram of the subalgebra $\mathfrak{so}(4)$ completes the periodic table by filling the 7th period to an inert gas \textbf{Og} (oganesson).}
\label{multiplet_n7}%
\end{figure}

It is easy to see that on each floor of the double $\SO(4,2)$-tower shown in Figure \ref{pic5}, the structure of Fock $(j,j)$-representations of the subgroup $\SO(4)$ (the integer part of the weight diagram of the subalgebra $\mathfrak{so}(4)$, see Figure 3 in \cite{Var2402}) is realized. The first floor ($n=1$) contains the first period, consisting of hydrogen \textbf{H} and helium \textbf{He}. \textbf{H} and \textbf{He} form a double singlet. On the floor $n=2$, the $m$-components (elements of the periodic table) form the $(1,1)$-diagram of the subalgebra $\mathfrak{so}(4)$ (see Figure \ref{multiplet_n2}), containing the second period: a double singlet (\textbf{Be},\textbf{Li}) and two triplets (\textbf{B},\textbf{C},\textbf{N}), (\textbf{Ne},\textbf{F},\textbf{O}).
Next, on the sheet $n=3$ (see Figure \ref{multiplet_n3}), we have a $(2,2)$-diagram of the subalgebra $\mathfrak{so}(4)$, which contains the 3rd period in full, as well as the transition metals of the side subgroups of the 4th period from scandium \textbf{Sc} to manganese \textbf{Mn} and from iron \textbf{Fe} to zinc \textbf{Zn}, forming two quintets.
Climbing higher to the next $n=4$ floor of the $\SO(4,2)$-tower, we come to the $(3,3)$-diagram of the subalgebra $\mathfrak{so}(4)$, which completes the filling of the 4th period with inert gas \textbf{Kr} (krypton), see Figure \ref{multiplet_n4}. $(3,3)$-diagram in Figure \ref{multiplet_n4} also contains transition metals of the 5th period from yttrium \textbf{Y} to technetium \textbf{Tc} and from ruthenium \textbf{Ru} to cadmium \textbf{Cd}. Here, within the framework of the $(3,3)$-diagram of Figure \ref{multiplet_n4}, the entire lanthanide family \textbf{La}, \textbf{Ce}, $\ldots$, \textbf{Yb} is located.
Rising even higher on the 5th floor, we come to another $(3,3)$-diagram of the subalgebra $\mathfrak{so}(4)$, see Figure \ref{multiplet_n5}, which, in turn, completes the filling of the 5th period with an inert gas \textbf{Xe} (xenon). $(3,3)$-diagram in Figure \ref{multiplet_n5} contains transition metals of the 6th period from lutetium \textbf{Lu} to rhenium \textbf{Re} and from osmium \textbf{Os} to mercury \textbf{Hg}. Here, within the framework of the $(3,3)$-diagram of Figure \ref{multiplet_n5}, the entire family of actinoids \textbf{Ac}, \textbf{Th}, $\ldots$, \textbf{No} is located.
Next, on the sheet $n=6$ (see Figure \ref{multiplet_n6}), we have the $(2,2)$-diagram of the subalgebra $\mathfrak{so}(4)$, which completes the filling of the 6th period with the inert gas \textbf{Rn} (radon). This diagram contains transition metals of the 7th period from lawrencium \textbf{Lr} to borium \textbf{Bh} and from hassium \textbf{Hs} to copernicium \textbf{Cn}.
Finally, on the floor $n=7$ (see Figure \ref{multiplet_n7}), we have the $(1,1)$-diagram of the subalgebra $\mathfrak{so}(4)$, which completes the filling of the 7th period with an inert gas \textbf{Og} (oganeson). The seventh period completes the filling of the periodic system.
The floor $n=8$ contains a double singlet of hypothetical elements \textbf{Uue} and \textbf{Ubn}, from which the construction of the Seaborg table begins and then the 10-periodic extension.

Thus, all elements of the periodic table are grouped into weight $\SO(4)$-diagrams of the subalgebra $\mathfrak{so}(4)$, located on eight floors of the weight diagram (double $\SO(4,2)$-tower, Figure \ref{pic5}) of the algebra $\mathfrak{so}(4,4)$.

On each floor of the left ($s=-1/2$) and right ($s=+1/2$) towers (see Figure \ref{pic3}) there are $n^2$ states, which corresponds to the dimension of representations of the subgroup $\SO(4)$, and on the floors of the combined tower (see Figure \ref{pic5}), we have implementations of weight diagrams of the subalgebra $\mathfrak{so}(4)$. Thus, the number of states (elements) on each floor of a combined $\SO(4,2)$-tower is determined by the Rydberg sequence $2n^2$:
\begin{equation}\label{Horizont}
2=2\cdot 1^2,\; 8=2\cdot 2^2,\; 18=2\cdot 3^2,\; 32=2\cdot 4^2,\;\ldots,
\end{equation}
 which Sommerfeld called ``cabalistic'' in his book \cite{Zommer}. Then there is \textit{period doubling} (except for the first one):
\begin{equation}\label{Vertical}
2,\;8,\;8,\;18,\;18,\;32,\;32,\;\ldots.
\end{equation}
The first doubling (\ref{Horizont}) is sometimes called ``horizontal'' (or \textit{spin}), the second (\ref{Vertical}) is ``vertical''.

\section{Seaborg Tower}
\begin{figure}[p] %
\centering
\includegraphics[width=16cm]{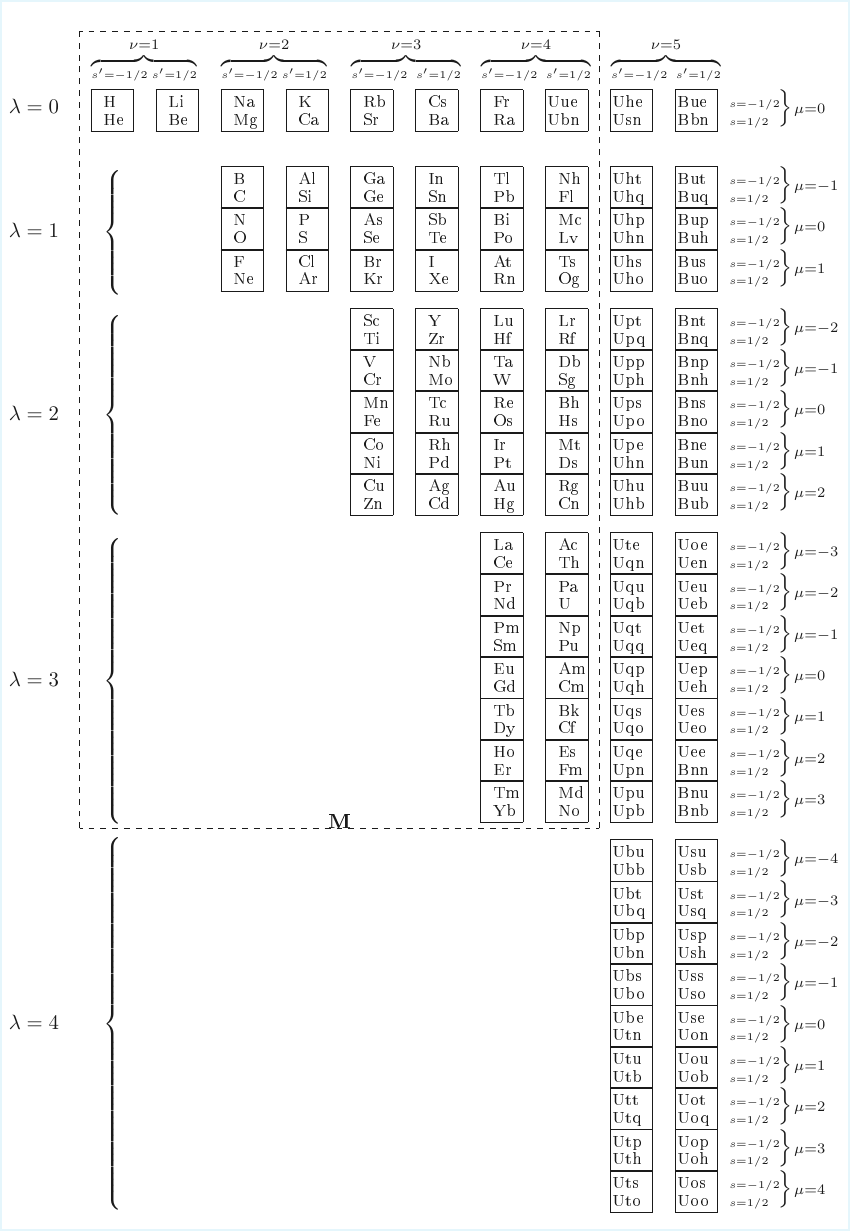}\\
\caption{The Seaborg table in the Janet-like form of the basic representation $F^+_{ss^\prime}$ of the Fet group. Quantum numbers $(\nu,\lambda,\mu,s,s^\prime)$ correspond to the first Fet basis.}
\label{SeaborgTable}%
\end{figure}

The Seaborg table is an eight-period extension of the periodic table. The eighth period begins with the elements Ununennium (eka-francium) with $Z=119$ and \textbf{Ubn} -- Unbinium (eka-radium) with $Z=120$. According to the Bohr model, the filling of the $g$ shell begins with the 121st element (the formation of the $g$ family). In the Fet \cite{Fet} model, the $g$-shell corresponds to the quantum numbers $\nu=5$ and $\lambda=4$ of the symmetry group $\SO(2,4)\otimes\SU(2)\otimes\SU(2)^\prime$ (see also \cite{Var1801}). Figure \ref{SeaborgTable} shows the Seaborg table in the form of the basic representation $F^+_{ss^\prime}$ of the Fet group for the basis
\begin{multline}
|\nu,\lambda,\mu,s,s^\prime\rangle,\quad \nu=1,2,\ldots;\; \lambda=0,1,\ldots, \nu-1;\\
\mu=-\lambda,-\lambda+1,\ldots,\lambda-1,\lambda;\;s=-1/2,1/2,\;s^\prime=-1/2,1/2.\label{Basis1}
\end{multline}
In turn, the basis (\ref{Basis1}) corresponds to the following chain of groups:
\begin{equation}
G\supset G_1\supset G_2\longmapsto
\SO(2,4)\otimes\SU(2)\otimes\SU(2)\supset\SO(4)\otimes\SU(2)\supset\SO(3)\otimes\SU(2),\label{Chain1}
\end{equation}
according to which the basic representation $F^+_{ss^\prime}$ of the Fet group is reduced by subgroups of the chain, i.e. the main multiplet is divided into smaller multiplets. The dotted frame with the symbol M in Figure \ref{SeaborgTable} highlights the periodic table. Ten more multiplets are added to the twenty multiplets of the periodic table as part of the eight-period extension (quantum numbers $\nu=5$, $\lambda=4$).

\begin{figure}[p] %
\centering
\includegraphics[width=10.5cm]{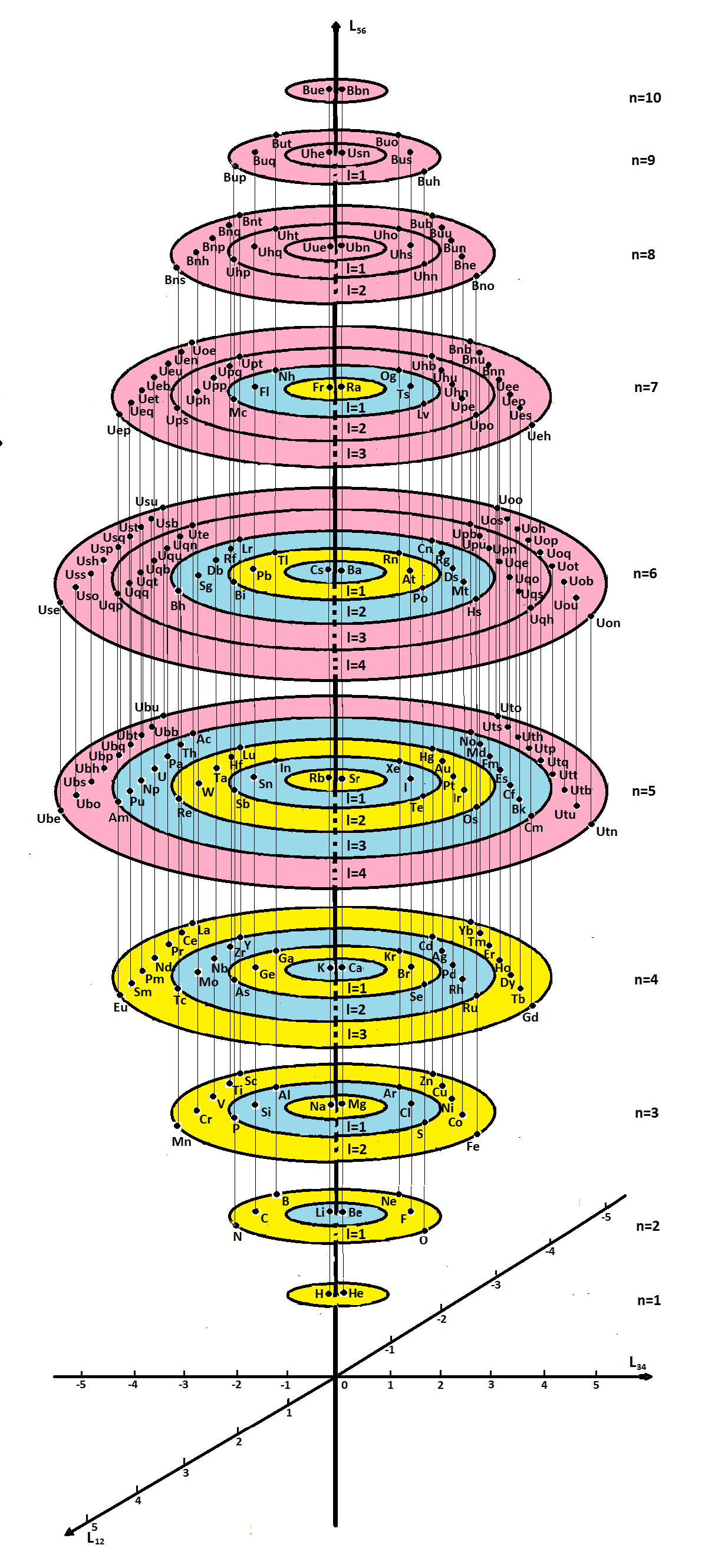}\\
\caption{\small Seaborg Tower (8-periodic extension) in the form of a combined weight diagram of the algebra $\mathfrak{so}(4,4)$. This double $\SO(4,2)$-tower is a superstructure above the Mendeleev Tower shown in Figure \ref{pic5}. The rings included in the superstructure are colored red.}
\label{pic_Seaborg}%
\end{figure}

Returning to the weight diagram of the algebra $\mathfrak{so}(4,4)$, we see that the eight-periodic extension of the periodic table can be represented as the following $\SO(4,2)$-tower, shown in Figure \ref{pic_Seaborg}. The elements included in the eight-period extension belong to rings colored red.
Unlike the Mendeleev Tower in Figure \ref{pic5}, the Seaborg Tower in Figure  \ref{pic_Seaborg} consists of 10 floors (levels) defined by the main quantum number $\nu\sim n$ (the eigenvalue of the radial generator $\bsL_{56}$), and is a kind of ``superstructure'' above the Mendeleev Tower, consisting of 7 periods.

So, the eighth period (eighth floor $n=8$ of $\SO(4,2)$-tower in Figure \ref{pic_Seaborg}) begins with the hypothetical elements \textbf{Uue} -- ununium (eka-francium) with $Z=119$ and \textbf{Ubn} -- unbinium (eka-radium) with $Z=120$, inhabiting the ring ($n=8,l=0$). This is followed by the transition \textbf{Ubn} $\rightarrow$ \textbf{Ubu} to the outer ring ($n=5, l=4$) of the fifth floor and filling this ring with elements of the superactinoid family of the 8th period \textbf{Ubu}. \textbf{Ubb}, $\ldots$, \textbf{Uto}, which includes 18 elements from $Z=121$ to $Z=138$. After lifting \textbf{Uto} $\rightarrow$ \textbf{Ute} on the 6th floor, the ring ($n=6,l=3$) is filled with elements of the transactinoid family of the 8th period \textbf{Ute}, \textbf{Uqn}, $\ldots$, \textbf{Upb} (homologues of the actinoid family of the ring ($n=5,l=3$) of the seventh period).The transactinoid family of the 8th period (as well as the actinoid family of the seventh period) contains 14 elements from $Z=139$ to $Z=152$. This is followed by lifting \textbf{Upb} $\rightarrow$ \textbf{Upt} to the floor $n=7$ and filling the ring ($n=7,l=2$) with transition metals \textbf{Upt}, \textbf{Upq}, $\ldots$, \textbf{Uhb} of the eighth period (homologues of transition metals \textbf{Lr}, \textbf{Rf}, $\ldots$, \textbf{Cn} of the seventh period). The number of transition metal elements of the 8th period is 10 (from $Z=153$ to $Z=162$). Climbing even higher, to the eighth floor, \textbf{Uhb} $\rightarrow$ \textbf{Uht}, we come to the settlement of the ring ($n=8,l=1$) of the $\SO(4,2)$-tower in Figure \ref{pic_Seaborg} by post-transition metals \textbf{Uht}, \textbf{Uhq}, $\ldots$, \textbf{Uho} of the eighth period (homologues of post-transition metals \textbf{Nh}, \textbf{Fl}, $\ldots$, \textbf{Og} of the seventh period). In this case, the element \textbf{Uho} (ungexocium, $Z=168$), which is a homologue of the oganesson \textbf{Og}, should be considered an inert gas. The number of transition metal elements of the 8th period is 6 (from $Z=163$ to $Z=168$). The settlement of the ring ($n=8,l=1$) with post-transition metals completes the filling of the eighth period. The total number of elements included in the eighth period is 50.

The ninth period (floor $n=9$), which is the double of the 8th period, begins with the elements \textbf{Uhe} (unhexennium, $Z=169$) and \textbf{Usn} (unseptnilium, $Z=170$) of the inner circle ($n=9,l=0$). This is followed by the descent \textbf{Usn} $\rightarrow$ \textbf{Usu} to the outer ring ($n=6,l=4$) of the sixth floor and filling this ring with elements of the superactinoid family of the 9th period \textbf{Usu}, \textbf{Usb}, $\ldots$, \textbf{Uoo} (18 elements from $Z=171$ to $Z=188$). The elements of this family are homologues of the elements of the superactinoid family of the 8th period, which is indicated in Figure \ref{pic_Seaborg} with vertical lines. After lifting \textbf{Uoo} $\rightarrow$ \textbf{Uoe} on the 7th floor, the ring ($n=7,l=3$) is populated with elements of the transactinoid family of the 9th period \textbf{Uoe}, \textbf{Uen}. $\ldots$, \textbf{Bnb} (homologues of the transactinoid family of the ring ($n=6,l=3$) of the 8th period). The transactinoid family of the 9th period, like the transactinoid family of the 8th period, contains 14 elements from $Z=189$ to $Z=202$. In Figure \ref{pic_Seaborg} vertical lines clearly show the homological chain of these families; lanthanides ($n=4,l=3$) $\rightarrow$ actinoids ($n=5,l=3$) $\rightarrow$ transactinoids of the 8th period ($n=6,l-3$) $\rightarrow$ transactinoids of the 9th period ($n=7,l=3$). This is followed by the rise \textbf{Bnb} $\rightarrow$ \textbf{Bnt} to the floor $n=8$ and filling the ring ($n=8,l=2$) with transition metals \textbf{Bnt}, \textbf{Bnq}, $\ldots$, \textbf{Bub} of the ninth period (homologues of transition metals of the eighth period). The number of transition metal elements of the 9th period is 10 (from $Z=203$ to $Z=212$). In Figure \ref{pic_Seaborg} the homologous chain of transition metals can be traced from $n=3$ to $n=8$: ($n=3,l=2$) $\rightarrow$ ($n=4,l=2$) $\rightarrow$ ($n=5,l=2$) $\rightarrow$ ($n=6,l=2$) $\rightarrow$ ($n=7,l=2$) $\rightarrow$ ($n=8,l=2$). Climbing even higher, to the ninth floor, \textbf{Bub} $\rightarrow$ \textbf{But}, we come to the settlement of the ring ($n=9,l=1$) with post-transition metals \textbf{But}, \textbf{Buq}, $\ldots$, \textbf{Buo} of the ninth period (homologues of the post-transition metals of the ring ($n=8,l=$1) of the eighth period). The number of ring elements ($n=9,l=1$) is 6 (from $Z=213$ to $Z=218$). Thus, the homological chain of rings of post-transition metals looks like this: ($n=2,l=1$) $\rightarrow$ ($n=3,l=1$) $\rightarrow$ ($n=4,l=1$) $\rightarrow$ ($n=5,l=1$) $\rightarrow$ ($n=6,l=1$) $\rightarrow$ ($n=7,l=1$) $\rightarrow$ ($n=8,l=1$) $\rightarrow$ ($n=9,l=1$). The element \textbf{Buo} (biunoctium, $Z=218$), which is a homologue of the elements \textbf{Uho} and \textbf{Og}, should be considered an inert gas on which the ninth period ends. The total number of elements included in the ninth period is 50. Thus, the Rydberg sequence of period lengths for the Seaborg Tower (Figure \ref{pic_Seaborg}) takes the form
\begin{equation}\label{Vertical2}
2,\;8,\;8,\;18,\;18,\;32,\;32,\;50,\;50\ldots.
\end{equation}

The tenth floor of the Seaborg Tower contains the elements \textbf{Bue} ($Z=219$) and \textbf{Bbn} ($Z=220$) in a circle ($n=10,l=0$), from which the tenth period begins and the further construction of the 10-periodic extension of the system of elements.

\section{10-Periodic Tower}
As noted above, in the Rumer-Fet-Barut approach, the periodic system of elements is understood as a single quantum system, where the Fet group $G_F$ (or the Ostrovsky group $G_O$) plays the role of dynamic symmetry linking the various states (chemical elements) of the quantum system. The symmetry $G$ of a quantum system can be represented as a quantum \textit{transition} between its states. It is natural to assume that the operators of the group $G$ or its subgroups connect related states. A chain of nested subgroups leads to a hierarchical classification of states.

The dynamic symmetry is defined by a chain of nested Lie groups:
\[
G=G_0\supset G_1\supset G_2\supset\ldots\supset G_k.
\]
\textit{The system} with this dynamic symmetry is given by an irreducible unitary representation $\fP$ of the group $G$ in the physical Hilbert space $\bsH_{\rm phys}$. The reduction $G/G_1$ of the representation $\fP$ of the group $G$ by the subgroup $G_1$ leads to the decomposition of $\fP$ into an orthogonal sum of irreducible representations $\fP^{(1)}_i$ of the subgroup $G_1$:
\[
\fP=\fP^{(1)}_1\oplus\fP^{(1)}_2\oplus\ldots\oplus\fP^{(1)}_i\oplus\ldots.
\]
In turn, the reduction $G_1/G_2$ of the representation of the group $G_1$ by the subgroup $G_2$ leads to the decomposition of the representations $\fP^{(1)}_i$ on irreducible representations $\fP^{(2)}_{ij}$ of the group $G_2$:
\[
\fP^{(1)}_i=\fP^{(2)}_{i1}\oplus\fP^{(2)}_{i2}\oplus\ldots\oplus\fP^{(2)}_{ij}\oplus\ldots
\]
etc.\footnote{For example, one of the basic supermultiplets of $\SU(3)$ theory (the baryon octet $F_{1/2}$), based on the eight-dimensional regular representation $\Sym^0_{(1,1)}$, admits the following $\SU(3)/\SU(2)$-reduction on isotopic multiplets of the subgroup $\SU(2)$: $\Sym^0_{(1,1)}=\Phi_3\oplus\Phi_2\oplus\overset{\ast}{\Phi}_2\oplus\Phi_0$, where $\Phi_3$ is a triplet, $\Phi_2$ and
$\overset{\ast}{\Phi}_2$ -- doublets, $\Phi_0$ -- singlet. Similarly, for the hypermultiplets of the $\SU(6)$ theory (56-plet baryons and 35-plet mesons), there are $\SU(6)/\SU(3)$- and $\SU(6)/\SU(4)$-reductions, where $\SU(4)$ is the Wigner subgroup \cite{Fet,Var15d}.}

The first Fet basis (\ref{Basis1}) corresponds to the chain of groups (\ref{Chain1}), according to which the basic representation $F^+_{ss^\prime}$ of the Fet group is reduced by subgroups of the chain, i.e. the main multiplet is divided into smaller multiplets. Ten more multiplets are added to the twenty multiplets of the periodic table as part of the eight-period extension (quantum numbers $\nu=5$, $\lambda=4$). The average masses of these multiplets are calculated according to the Fet formula
\begin{equation}\label{Mass1}
m=m_0+a\left[s^\prime(2\nu-3)-5\nu+\frac{11}{2}+2(\nu^2-1)\right]-b\cdot\lambda(\lambda+1),
\end{equation}
where $m_0$, $a$, $b$ are coefficients that cannot be deduced from theory. The mass formula (\ref{Mass1}) corresponds to a chain of groups (\ref{Chain1}). The formula (\ref{Mass1}) is similar to the ``first perturbation'' in the $\SU(3)$- and $\SU(6)$-theories, which allows us to calculate the average mass of the multiplet elements\footnote{So, in the $\SU(3)$-theory, the Gell-Mann-Okubo mass formula holds
\[
m=m_0+\alpha+\beta Y+\gamma\left[I(I+1)-\frac{1}{4}Y^2\right]
+\alpha^\prime-\beta^\prime
Q+\gamma^\prime\left[U(U+1)-\frac{1}{4}Q^2\right],
\]
in which, according to $\SU(3)/\SU(2)$-reduction, quantum numbers ($I$ - isotopic spin, $Y$ - hypercharge) standing in the first square bracket, define the ``first perturbation'', which leads to the so-called \textit{hypercharge} splitting of masses, i.e. splitting the multiplet of the group $\SU(3)$ into smaller multiplets of the subgroup $\SU(2)$. ``The second perturbation'' is given by the quantum numbers in the second square bracket ($Q$ is the charge, $U$ is the isotopic spin, which, unlike $I$, corresponds to a different choice of basis in the subgroup $\SU(2)$), which in turn leads to \textit{charge} (element-wise) splitting of masses inside multiplets of $\SU(2)$.}. To obtain an analog of the ``second perturbation'' leading to mass splitting inside the multiplets of the group $G_2=\SO(3)\otimes\SU(2)$, it is necessary to find a further extension of the chain of groups $G\supset G_1\supset G_2$ (\ref{Chain1}), i.e. the problem it boils down to finding another subgroup of $G_3$. Then $G_2/G_3$-reduction will give an element-wise splitting of the masses.

\begin{figure}[p] %
\centering
\includegraphics[width=15cm]{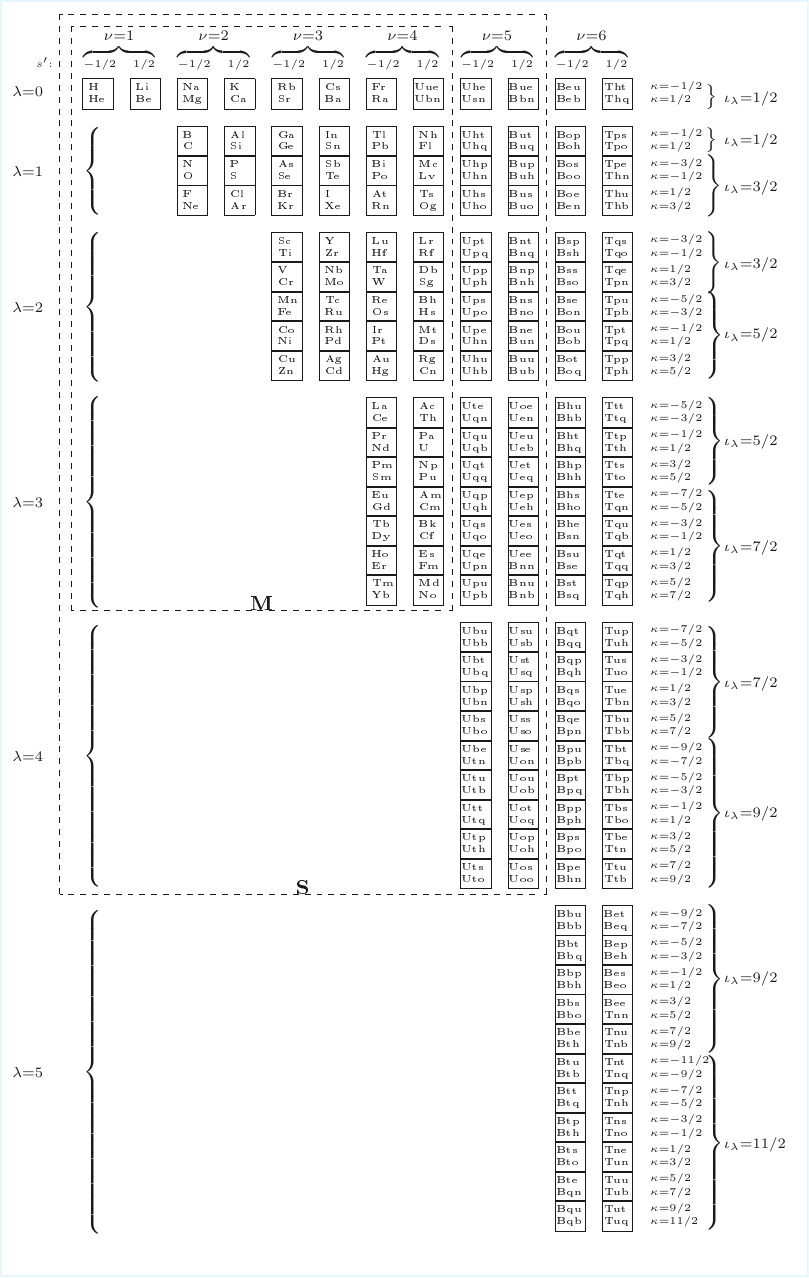}\\
\caption{10-periodic extension of the periodic table in the Janet-like form of the basic representation $F^+_{ss^\prime}$ of the Fet group. Quantum numbers $(\nu,s^\prime,\lambda,\iota_\lambda,\kappa)$ correspond to the second Fet basis.}
\label{pic_TenPeriodic}
\end{figure}

$G_2/G_3$-reduction leads to the following (elongated) chain of groups:
\begin{multline}
G\supset G_1\supset G_2\supset G_3\longmapsto\\
\SO(2,4)\otimes\SU(2)\otimes\SU(2)\supset\SO(4)\otimes\SU(2)\supset\SO(3)\otimes\SU(2)\supset\SO(3)_c.\label{Chain2}
\end{multline}
Lengthening the chain of groups requires the introduction of a new basis, the vectors of which must belong to the smallest multiplets of the symmetry group, i.e., the multiplets of the subgroup $G_3$. The vectors $|\nu,\lambda,\mu,s,s^\prime\rangle$ of the basis (\ref{Basis1}) corresponding to the chain of groups (\ref{Chain1}) no longer constitute a dedicated basis, since $\mu$, $s$ are no longer the quantum numbers of the symmetry group, i.e. these vectors do not belong to the spaces of irreducible representations of the group $G_3$. The new basis is defined as follows. Since $\nu$, $s^\prime$, $\lambda$ are associated with the groups $G$, $G_1$, $G_2$, they remain the quantum numbers of the chain (\ref{Chain1}), and new quantum numbers are introduced instead of $\mu$, $s$, related to $G_3$. One of them is $\iota_\lambda$, associated with the Casimir operator of the subgroup $G_3$, equal to $\sum^3_{k=1}(\boldsymbol{\tau}_k+\bsJ_k+\bsK_k)^2$. In this case, the two multiplets $G_3$ into which the representations of the group $G_2$ are divided correspond to $\iota_\lambda=\lambda-1/2$ and $\iota_\lambda=\lambda+1/2$, whence $2\lambda=2\iota_\lambda+1$, $2\lambda+2=2\iota_\lambda+1$. Another quantum number $\kappa$ is the eigenvalue of the operator $q_3=\boldsymbol{\tau}_3+\bsJ_3+\bsK_3$ belonging to the Lie algebra of the group $G_3=\SO(3)_c$. Thus, the new basis corresponding to the chain of groups (\ref{Chain2}) has the form
\begin{multline}
|\nu,s^\prime,\lambda,\iota_\lambda,\kappa\rangle,\quad \nu=1,2,\ldots;\;s^\prime=-1/2,1/2;\;\lambda=0,1,\ldots, \nu-1;\\
\iota_\lambda=\lambda-1/2,\lambda+1/2;\;\kappa=-\iota_\lambda, -\iota_\lambda+1,\ldots,\iota_\lambda-1,\iota_\lambda.\label{Basis2}
\end{multline}
The ten-period table in Figure \ref{pic_TenPeriodic} corresponds to the reduction chain \ref{Chain2}.
The chain of groups $G\supset G_1\supset G_2\supset G_3$ (\ref{Chain2}) allows us to obtain an element-wise splitting of the masses of the basic representation $F^+_{ss^\prime}$ of the Fet group. For this purpose, we use the mass formula introduced in \cite{Var1802}:
\begin{multline}
m=m_0+a\left[s^\prime(2\nu-3)-5\nu+\frac{11}{2}+2(\nu^2-1)\right]-b\cdot\lambda(\lambda+1)+\\
+a^\prime\left[2\kappa-0,1666\kappa^3+0,0083\kappa^5-0,0001\kappa^7\right]+\left(b^\prime\iota_\lambda\right)^p-1,
\label{Mass2}
\end{multline}
where
\[
p=\left\{\begin{array}{rl}
0, & \mbox{if $\iota_\lambda=\lambda-1/2$};\\
1, & \mbox{if $\iota_\lambda=\lambda+1/2$}.
\end{array}\right.
\]
The theoretical masses of the elements of the periodic table are calculated according to the mass formula (\ref{Mass2}) for the values $m_0=1$, $a=17$, $b=5.5$, $a^\prime=1.15$, $b^\prime=3.9$. It is shown \cite{VPB22} that the masses of the elements determined by the formula (\ref{Mass2}) are more accurate the heavier the elements, therefore, the formula (\ref{Mass2}) is asymptotic \cite{VPB22}.

\begin{figure}[p] %
\centering
\includegraphics[width=9cm]{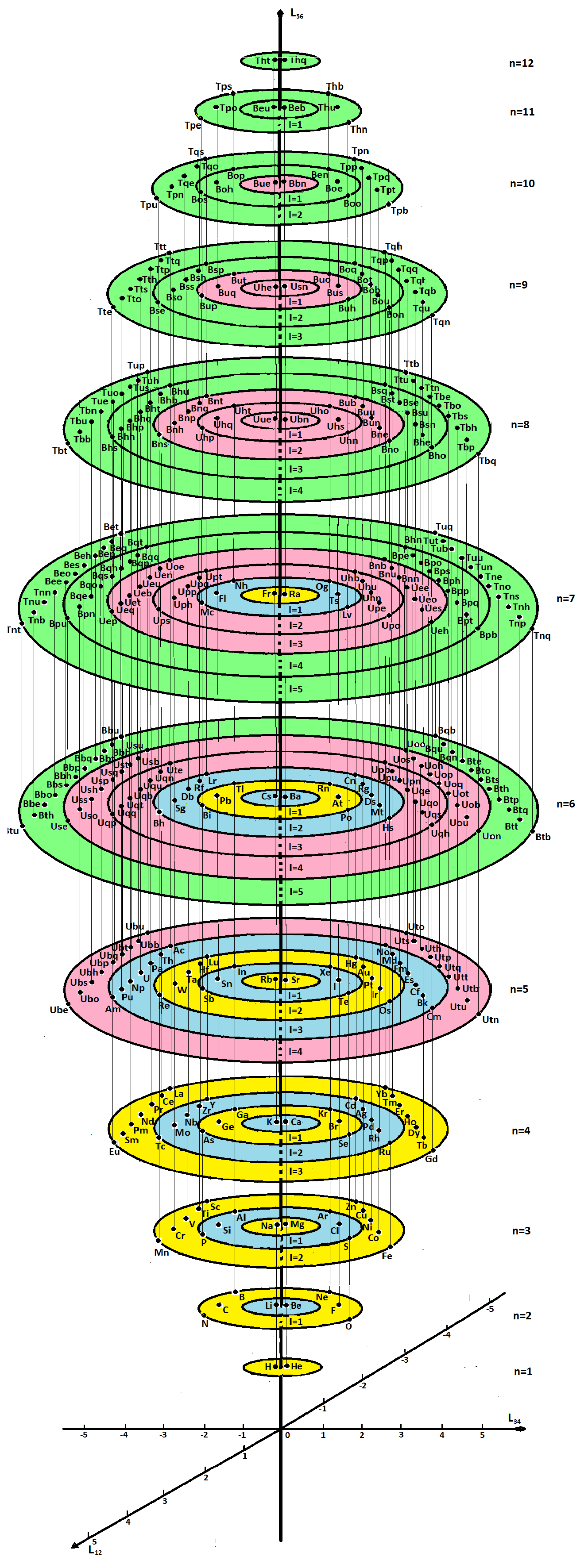}\\
\vspace{-0.2cm}
\caption{\small 10-periodic Tower in the form of a combined weight diagram of the algebra $\mathfrak{so}(4,4)$. This double $\SO(4,2)$-tower is a superstructure above the Seaborg Tower shown in Figure \ref{pic_Seaborg}. The rings included in the superstructure are colored green.}
\label{pic_10_Tower}
\end{figure}

The ten-period $\SO(4,2)$-tower (see Figure \ref{pic_10_Tower}), corresponding to the table \ref{pic_TenPeriodic}, is the next superstructure above the Seaborg Tower (see Figure \ref{pic_Seaborg}). The elements included in the ten-period extension belong to rings colored green. Unlike the Seaborg Tower in Figure \ref{pic_Seaborg}, the ten-period tower in Figure \ref{pic_10_Tower} consists of 12 floors (levels) defined by the quantum number $\nu\sim n$ (the eigenvalue of the radial generator $\bsL_{56}$).

The tenth period (floor $n=10$ of the $\SO(4,2)$-tower in Figure \ref{pic_10_Tower}) begins with the elements \textbf{Bue} (biunennium, $Z=219$) and \textbf{Bbn} (bibinilium, $Z=220$), inhabiting the inner circle ($n=10,l=0$). This is followed by the descent \textbf{Bbn} $\rightarrow$ \textbf{Bbu} to the outer ring ($n=6,l=5$) of the sixth floor and filling this ring with elements of the hyperactinoid family of the 10th period \textbf{Bbu}, \textbf{Bbb}, $\ldots$, \textbf{Bqb}, which includes 22 elements from $Z=221$ to $Z+242$. After lifting \textbf{Bqb} $\rightarrow$ \textbf{Bqt} on the 7th floor, we come to fill the ring ($n=7,l=4$) with elements of the superactinoid family of the 10th period \textbf{Bqt}, \textbf{Bqq}, $\ldots$, \textbf{Bhn}, which includes 18 elements from $Z=243$ to $Z=260$. The superactinoid family of the 10th period is a homologue of the superactinoid family of the 9th period \textbf{Usu}, \textbf{Usb}, $\ldots$, \textbf{Uoo}, which is indicated in Figure \ref{pic_10_Tower} with vertical lines. This is followed by the rise \textbf{Bhn} $\rightarrow$ \textbf{Bhu} to the floor $n=8$ and the settlement of the ring ($n=8,l=3$) with elements of the transactinoid family of the 10th period \textbf{Bhu}, \textbf{Bhb}, $\ldots$, \textbf{Bsq} (homologues of the transactinoid family 9-th period \textbf{Uoe}, \textbf{Uen}, $\ldots$, \textbf{Bnb}). The transactinoid family of the 10th period contains 14 elements from $Z=261$ to $Z=274$. Climbing even higher, to the ninth floor, \textbf{Bsq} $\rightarrow$ \textbf{Bsp}, we arrive at the settlement of the ring ($n=9,l=2$) with transition metals \textbf{Bsp}, \textbf{Bsh}, $\ldots$, \textbf{Boq} of the 10th period (homologues of transition metals \textbf{Bnt}, \textbf{Bnq}, $\ldots$, \textbf{Bub} of the 9th period). The number of transition metal elements of the 10th period is 10 (from $Z=275$ to $Z=284$). Climb to the tenth floor, \textbf{Boq} $\rightarrow$ \textbf{Bop}, leads to the filling of the ring ($n=10,l=1$) with post-transition metals \textbf{Bop}, \textbf{Boh}, $\ldots$, \textbf{Ben} of the tenth period (homologues of post-transition metals of the ninth period \textbf{But}, \textbf{Buq}, $\ldots$, \textbf{Buo}). The number of ring elements ($n=10,l=1$) is 6 (from $Z=284$ to $Z=290$). The settlement of the ring ($n=10,l=1$) with post-transition metals completes the filling of the tenth period. The total number of elements included in the tenth period is 72.

The eleventh period, which is the double of the 10th period, begins with the elements \textbf{Beu} (biennunium, $Z=291$) and \textbf{Beb} (biennbium, $Z=292$) inhabiting the inner circle ($n=11,l=0$). This is followed by the descent \textbf{Beb} $\rightarrow$ \textbf{Bet} to the outer ring ($n=7,l=5$) of the seventh floor and filling this ring with elements of the hyperactinoid family of the 11th period \textbf{Bet}, \textbf{Beq}, $\ldots$, \textbf{Tuq}, which includes 22 elements from $Z=293$ to $Z=314$ (homologues of the hyperactinoid family of the 10th period \textbf{Bbu}, \textbf{Bbb}, $\ldots$, \textbf{Bqb}). After lifting \textbf{Tuq} $\rightarrow$ \textbf{Tup} on the 8th floor, the ring ($n=8,l=4$) is filled with elements of the superactinoid family of the 11th period \textbf{Tup}, \textbf{Tuh}, $\ldots$, \textbf{Ttb}, which includes 18 elements from $Z=315$ to $Z=332$. This ring is a homologue of the superactinoid ring ($n=7,l=4$) of the 10th period. Climbing even higher, to the 9th floor \textbf{Ttb} $\rightarrow$ \textbf{Ttt}, we come to populate the ring ($n=9,l=3$) with elements of the transactinoid family of the 11th period \textbf{Ttt}, \textbf{Ttq}, $\ldots$, \textbf{Tqh}, which includes 14 elements from $Z=333$ to $Z=346$. This ring is a homologue of the transactinoid ring ($n=8,l=3$) of the 10th period, as shown in Figure \ref{pic_10_Tower} by vertical lines. Climb to the tenth floor, \textbf{Tqh} $\rightarrow$ \textbf{Tqs}, leads to the colonization of the ring ($n=10,l=2$) with elements of the transition metal family \textbf{Tqs}, \textbf{Tqo}, $\ldots$, \textbf{Tph} of the 11th period, which includes 10 elements from $Z=$347 to $Z=$356. In turn, this ring is a homologue of the ring ($n=9,l=2$) of transition metals of the 10th period. Finally, the rise \textbf{Tph} $\rightarrow$ \textbf{Tps} on the 11th floor leads to the settlement of the ring ($n=11,l=1$) with post-transition metals \textbf{Tps}, \textbf{Tpo}, $\ldots$, \textbf{Thb} of the 11th period. The number of ring elements ($n=11,l=1$) is 6 (from $Z=357$ to $Z=362$). The settlement of the ring ($n=11,l=1$) with post-transition metals completes the filling of the eleventh period. The total number of elements included in the 11th period is 72. Thus, the Rydberg sequence of period lengths for a 10-period tower looks like this:
\begin{equation}\label{Vertical3}
2,\;8,\;8,\;18,\;18,\;32,\;32,\;50,\;50,\;72,\;72\ldots.
\end{equation}

The ten-period tower shown in Figure \ref{pic_10_Tower} contains the following homologous chains:\\
1) \textbf{Alkaline and alkaline-earth metals}
\begin{multline}
(n=1,l=0)\rightarrow(n=2,l=0)\rightarrow(n=3,l=0)\rightarrow(n=4,l=0)\rightarrow
(n=5,l=0)\rightarrow\\
(n=6,l=0)\rightarrow(n=7,l=0)\rightarrow(n=8,l=0)\rightarrow(n=9,l=0)\rightarrow\\
(n=10,l=0)\rightarrow(n=11,l=0).
\end{multline}
2) \textbf{Post-transition metals}
\begin{multline}
(n=2,l=1)\rightarrow(n=3,l=1)\rightarrow(n=4,l=1)\rightarrow(n=5,l=1)\rightarrow\\
(n=6,l=1)\rightarrow(n=7,l=1)\rightarrow(n=8,l=1)\rightarrow(n=9,l=1)\rightarrow\\
(n=10,l=1)\rightarrow(n=11,l=1).
\end{multline}
3) \textbf{Transition metals}
\begin{multline}
(n=3,l=2)\rightarrow(n=4,l=2)\rightarrow(n=5,l=2)\rightarrow(n=6,l=2)\rightarrow
(n=7,l=2)\rightarrow\\
(n=8,l=2)\rightarrow(n=9,l=2)\rightarrow(n=10,l=2).
\end{multline}
4) \textbf{Lanthanides, actinoids, transactinoids}
\begin{multline}
(n=4,l=3)\rightarrow(n=5,l=3)\rightarrow(n=6,l=3)\rightarrow(n=7,l=3)\rightarrow\\
(n=8,l=3)\rightarrow(n=9,l=3).
\end{multline}
5) \textbf{Superactinoids}
\begin{equation}
(n=5,l=4)\rightarrow(n=6,l=4)\rightarrow(n=7,l=4)\rightarrow(n=8,l=4).
\end{equation}
6) \textbf{Hyperactinoids}
\begin{equation}
(n=6,l=5)\rightarrow(n=7,l=5).
\end{equation}

\section{Spin and Period Doubling}
P.-O. L\"{o}wdin notes in the article \cite{Low69} that the lack of a theoretical explanation for the period doubling (as stated in \cite{Scerri,Tyss}, which has been taking place so far) is equivalent to the lack of a theoretical understanding of the periodic table of chemical elements as a whole.

\begin{figure}[ht]
\centering
\vspace{-0.5cm}
\includegraphics[width=\textwidth]{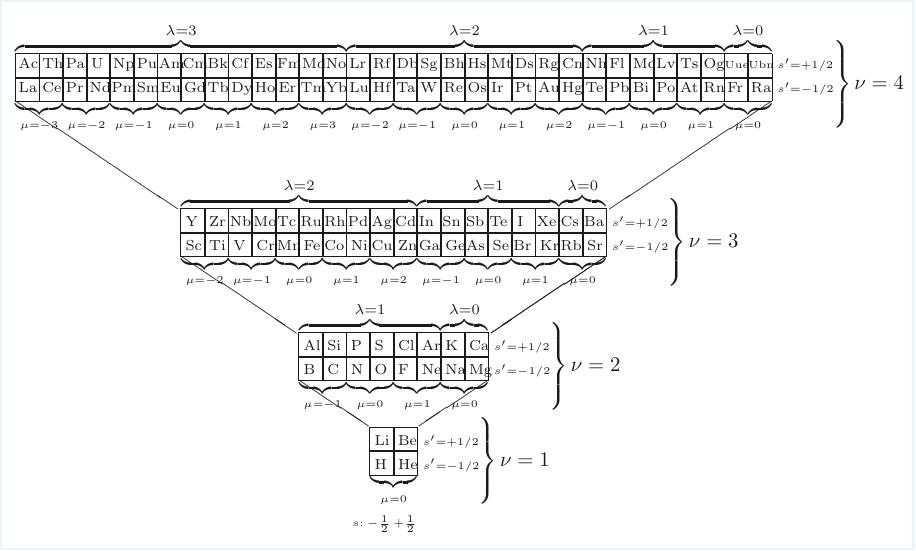}\\
\caption{The Janet table in a pyramidal form with quantum numbers of the Fet group (the first Fet basis).}
\label{pic4}
\end{figure}

Let's consider how the period doubling was described by Ostrovsky \cite{Ost81} and Fet \cite{Fet}\footnote{Earlier in these papers, Barut \cite{Bar72} tried to explain the period doubling by reducing representations of the conformal group $\SO(4,2)$ relative to the subgroup $\SO(3,2)$ (the anti-de Sitter group). O. Novaro \cite{Nov89} tried to explain the period doubling by means of the difference between two types of representations $(j,j,0)$ and $(j,0,j)$ of the Novaro-Berrondo group $G_{NB}=\SU(2)\otimes\SU(2)\otimes\SU(2)$.}. Ostrovsky group \cite{Ost81}
\begin{equation}\label{G_O}
G_O=\GO(4,2)\otimes\SU(2)_S\otimes\SU(2)_T
\end{equation}
and Fet group \cite{Fet}
\begin{equation}\label{G_F}
G_F=\SO(4,2)\otimes\SU(2)\otimes\SU(2)^\prime
\end{equation}
have a similar structure\footnote{In earlier works \cite{Fet79,Fet80} Fet interprets the period doubling by including the cyclic group $\dZ_2$: $\GO(4,2)\otimes\SU(2)\otimes\dZ_2$.}. The subgroup $\GO(4)\otimes\SU(2)_S\otimes\SU(2)_T$ in (\ref{G_O}) contains the symmetry $\GO(4)$, which leads to representations of dimension $n^2$. By extending the group $\GO(4)$ to $\GO(4)\otimes\SU(2)_S$, the dimensions of the representations are doubled to $2n^2$. The subscript $S$ here indicates the physical origin of the group $\SU(2)$ from the electron spin $m_s=\pm 1/2$. Ostrovsky called this ``horizontal'' doubling of period lengths \textit{spin doubling} (see (\ref{Horizont})). Fet calls this \textit{first doubling} $\SO(4,2)\otimes\SU(2)$, the representation of this group is $F^+_s=\varphi_2\otimes F^+$, where $\varphi_2$ is the unitary representation of the group $\SU(2)$ in the space $C(2)$, $F^+$ is an extension of the Fock representation  $F$ for the subgroup $\SO(4)$ to the conformal group $\SO(4,2)$. This leads to \textit{two copies} of the representations of the group $\SO(4,2)\otimes\boldsymbol{1}$, which are implemented in two different Hilbert spaces $\cH_+$ and $\cH_-$. Next, Fet introduces three generators $\boldsymbol{\tau}_+$, $\boldsymbol{\tau}_-$ and $\boldsymbol{\tau}_3$ of the algebra $\mathfrak{su}(2)$, where $\boldsymbol{\tau}_3$ acts as the Cartan generator, which is permuted with all generators of the subgroup $\SO(4,2)\otimes\boldsymbol{1}$ and distinguishes states from both subspaces $\cH_+$ and $\cH_-$, and the ladder operators $\boldsymbol{\tau}_\pm$ (Weyl generators) act as shift operators between $\cH_+$ and $\cH_-$. Figure \ref{pic4} shows the Janet table in a pyramidal form, where the cells of the table (chemical elements) are accompanied by the quantum numbers of the Fet group (the first Fet basis). It can be seen from the figure that the horizontal (spin) doubling corresponds to the double splitting of the $\mu$-components of the $\lambda$-multiplets.
``Vertical'' doubling of period lengths, known as the actual \textit{period doubling} (see (\ref{Vertical})) in the periodic table, was formulated by Ostrovsky in group-theoretic form by introducing a second group $\SU(2)$, denoted by $\SU(2)_T$ and formally analogous to the isospin group. Fet calls this \textit{second doubling} $\SO(4,2)\otimes\SU(2)\otimes\SU(2)^\prime$ (Fet group $G_F$), which is introduced similarly to the first doubling, i.e. the representation of this group has the form $F^+_{ss^\prime}=\varphi^\prime_2\otimes F^+_s$, where $\varphi^\prime_2$ is the unitary representation of the group $\SU(2)^\prime$ in the space $C(2)$. In turn, this leads to \textit{two copies} of the representations of the group $\SO(4,2)\otimes\SU(2)\otimes\boldsymbol{1}$, which are implemented in two different Hilbert spaces $\cH^\prime_+$ and $\cH^\prime_-$. Further, Fet introduced three generators $\boldsymbol{\tau}^\prime_+$, $\boldsymbol{\tau}^\prime_-$ and $\boldsymbol{\tau}^\prime_3$ of the algebra $\mathfrak{su}(2)^\prime$, where $\boldsymbol{\tau}^\prime_3$ acts as a Cartan generator that commutes with all generators of the subgroup $\SO(4,2)\otimes\SU(2)\otimes\boldsymbol{1}$ and distinguishes states from both subspaces $\cH^\prime_+$ and $\cH^\prime_-$, and the Weyl generators $\boldsymbol{\tau}^\prime_\pm$ act as shift (ladder) operators between $\cH^\prime_+$ and $\cH^\prime_-$. In Figure \ref{pic4}, the vertical doubling given by \textit{the fifth quantum number} $s^\prime$ corresponds to the double (vertical) splitting of the $\nu$ blocks (tiers of the Janet pyramid).

The main disadvantage of the Ostrovsky-Fet scheme of the group-theoretic description of period doubling is the artificial nature of the introduction of a fifth quantum number $s^\prime$, which has no real analogue, since all states (elements) of the periodic system are described by the four quantum numbers $(n,l,m,s)$. According to Fet \cite{Fet}, the set $(\nu,\lambda,\mu,s,s^\prime)$ of the five quantum numbers of the group $G_F$ defines all the states of the periodic table, while the quantum number $\nu$ is equal to $\nu=1/2(n+l+1)$ for an odd value of the sum of $n+l$ and $\nu=1/2(n+l)$ for an even value of $n+l$, which leads to a change in the Madelung rule.

Turning to the analysis of period doubling within the framework of $\SO(4,4)$, first of all it should be noted that the group $\SO(4,4)$ has a higher symmetry than tensor products (\ref{G_O}) and (\ref{G_F}), which are subgroups of rotation groups of ten-dimensional spaces. Accordingly, their Lie algebras have fifth rank. The Lie algebra $\mathfrak{so}(4,4)$ contains 28 independent generators (see section 3), while$\mathfrak{so}(4,2)\otimes\mathfrak{su}(2)\otimes\mathfrak{su}(2)^\prime$ has 21 generator. As noted above in section 3, the fourth generator $\bsL_{78}$ of the Cartan subalgebra $\fK\subset\mathfrak{so}(4,4)$, understood as a spin generator, commutes with all 15 generators of the subalgebra $\mathfrak{so}(4,2)$, which leads to a double splitting (\ref{IB1}) and (\ref{IB2}). The weight diagrams corresponding to the bases (\ref{IB1}) and (\ref{IB2}) are shown in Figure \ref{pic3}. In this case, the horizontal (spin) doubling of the Ostrovsky-Fet scheme is represented by two towers of representations: $s=-1/2$ (hydrogen line) and $s=1/2$ (helium line). The ``vertical'' (actual) period doubling is given by the floors (Haenzel circles) $K$, $L$, $M$, $N$, $O$, $P$, $Q$, $R$ of the corresponding $\SO(4,2)$-towers (pyramid tiers representations of the conformal group, see Figure 7 in \cite{Var2501}). At the same time, unlike the Ostrovsky-Fet scheme, it does not require the introduction of an additional fifth quantum number.

Within the framework of the proposed scheme for describing the periodic table, the fourth generator $\bsL_{78}$ of the Cartan subalgebra of the algebra $\mathfrak{so}(4,4)$ combines the functions of both generators $\boldsymbol{\tau}_3$ and $\boldsymbol{\tau}^\prime_3$ of the Ostrovsky-Fet scheme by virtue of higher symmetry of the group $\SO(4,4)$. Switching the generator $\bsL_{78}$ with all 15 generators of the subalgebra $\mathfrak{so}(4,2)$ leads to splitting of the Cartan-Weyl basis of the algebra $\mathfrak{so}(4,4)$ and, accordingly, to the first doubling (\ref{Horizont}). The weight diagram of the algebra $\mathfrak{so}(4,4)$ is four-dimensional, the axis of the generator $\bsL_{78}$ is perpendicular to the axis of the radial generator $\bsL_{56}$, as well as to the axes of the generators $\bsL_{12}$ and $\bsL_{34}$, i.e. the axis of the generator $\bsL_{78}$ is perpendicular to the three-dimensional weight space, which is based on the axes of the generators $\bsL_{12}$, $\bsL_{34}$, $\bsL_{56}$, which in turn leads to ``vertical'' (actual) period doubling (\ref{Vertical}). This eliminates the need to introduce a fifth quantum number.

Thus, \textit{the actual period doubling (\ref{Vertical}) is the result of the action of the fourth generator $\bsL_{78}$ (spin generator) of the Cartan subalgebra $\fK$ of the Lie algebra $\mathfrak{so}(4,4)$}. In this case, spin is the fourth degree of freedom, which has no analogue in the framework of the classical three-dimensional system. \textit{Spin is an effect of the fourth dimension}. That substantial three-dimensional image of an atom formed by means of recording equipment (mass spectrometers, electron and tunneling microscopes) is nothing more than the tip of an iceberg, the main part of which is immersed in the world of four (and maybe more) dimensions. Heisenberg said: ``Atoms are not things''.

\section{Antimatter}
\begin{figure}[p] %
\centering
\includegraphics[width=16cm]{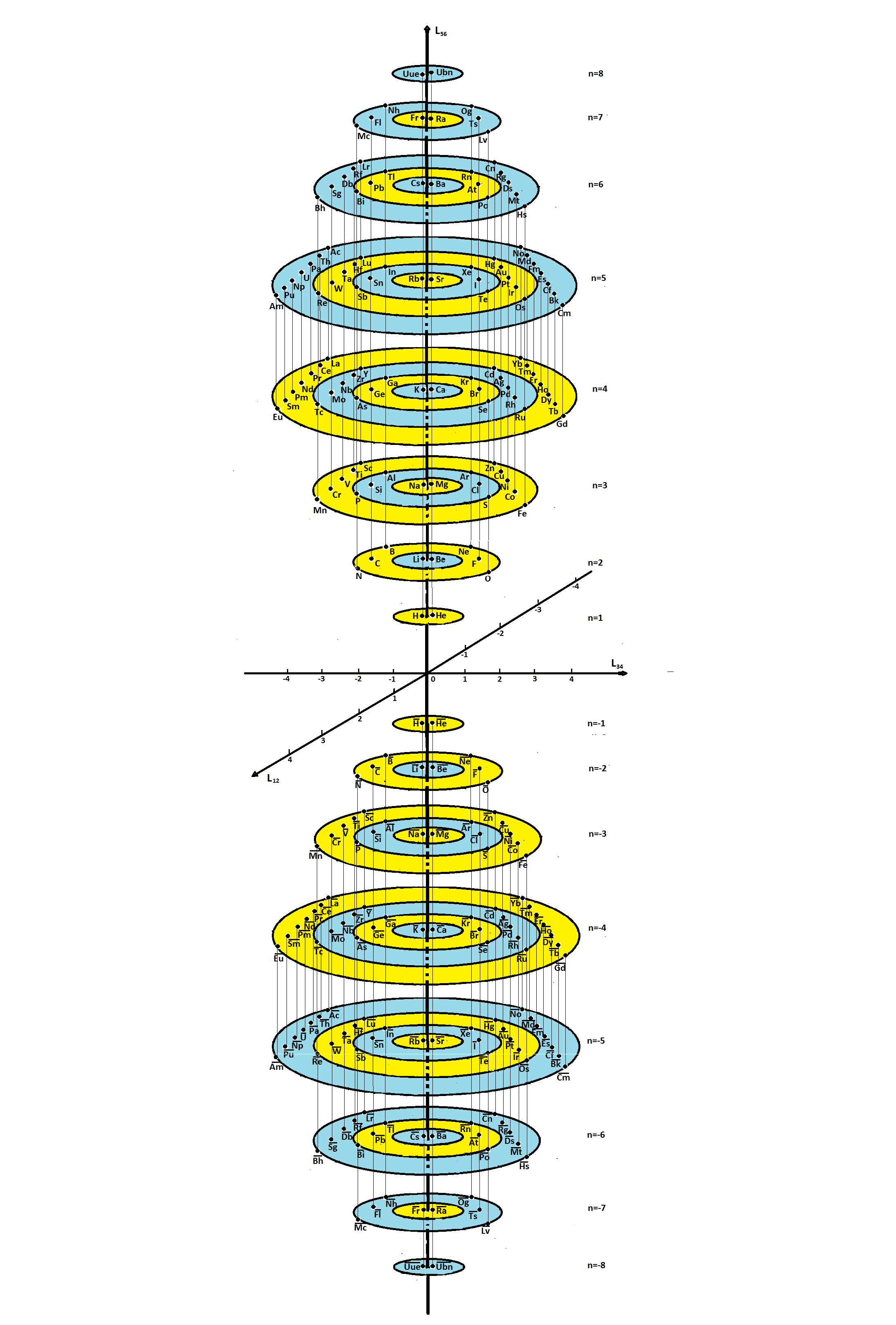}\\
\caption{ The elements ($n=1,2,\ldots$) and antielements ($n=-1,-2,\ldots$) of the periodic table in the representation of the doubled three-dimensional projection of the weight diagram of the Lie algebra $\mathfrak{so}(4,4)$.}
\label{pic_MA}%
\end{figure}
The first evidence of the existence of \textit{antihydrogen} was obtained at accelerators in the 1990s. The capture of antihydrogen atoms in a trap was first demonstrated by the Antihydrogen Laser Physics Apparatus (ALPHA) group at CERN in 2010. It is logical to assume that the existence of antihydrogen atoms entails the existence of atoms \textit{antihelium}, \textit{anti-lithium}, etc., i.e. the existence of a dual Mendeleev \textit{anti-table}. The group-theoretic description naturally includes the consideration of the \textit{antielements} of the periodic system already at the level of the $\SO(4,2)$ subgroup. The weight diagram of the algebra $\mathfrak{so}(4,2)$ consists of two pyramids symmetrical with respect to the plane formed by the Cartan generators $\sL_3$ and $\sA_3$ (see Figure 6 in \cite{Var2402}). In this case, the eigenvalue of the radial generator $\Delta_3$ for the lower pyramid takes negative values. Thus, the inclusion of antimatter in the general group-theoretic scheme for describing a periodic system requires the introduction of negative values of the main quantum number $n$, i.e., the eigenvalues of the radial generator $\bsL_{56}$. The limitation to only positive values of the number $n$ in the Bohr model is due to the geometric characteristics of the Keplerian orbits of the Rutherford planetary model (for more details, see \cite{Zommer}). It should be noted that in the Bohr model, the quantum numbers $n$, $l$, and $m$ correspond to the radial and orbital motions of the classical system (Kepler problem) describing the motions of the ``constituent parts'' of an atom. At the same time, the fourth quantum number $s$ has no classical analogue, which once again indicates the palliative nature of this model. In the article \cite{Bar72} on the group structure of the periodic table within the conformal group $\SO(4,2)$, A. Barut introduces the quantum numbers $\nu$, $\lambda$, $\mu_\lambda$ of the group $\SO(4,2)$ in order to consider chemical elements as different \textit{states} of a single \textit{quantum system}. In order not to be biased against the quantum numbers $n$, $l$, $m_l$ of Bohr's mechanical (planetary) model describing hydrogen-like systems, Barut intentionally introduces the symbols $\nu$, $\lambda$, $\mu_\lambda$, which have, first of all, group-theoretic (non-mechanical) meaning, although Barut leaves the range of their variation the same as that of the ``hydrogen'' quantum numbers of the Bohr model. Obviously, this limitation of the range of variation of the numbers $\nu$, $\lambda$, $\mu_\lambda$, which are the eigenvalues of the Cartan generators of the algebra $\mathfrak{so}(4,2)$, is an artificial assumption resulting from the requirement of agreement with the Bohr model. Ostrovsky similarly distinguishes between ordinary (``hydrogen'') quantum numbers and abstract $\SO(4,2)$ symbols, denoting the latter with a tilde sign: $\tilde{n}$, $\tilde{l}$, $\tilde{m}_l$ \cite{Ost96}. Fet in the book \cite{Fet} adheres to the symbolism of Barut. Following Barut, Ostrovsky and Fet limit the range of variation of the eigenvalue $\nu$ of the radial generator to positive values.

In turn, for the algebra $\mathfrak{so}(4,4)$, the weight space is defined by four Cartan generators $\bsL_{56}=\Delta_3$, $\bsL_{12}=\sL_3$, $\bsL_{34}=\sA_3$ and $\bsL_{78}$, which serve as the basis of a four-dimensional orthogonal coordinate system. Let's define the ket-vector
\begin{equation}\label{Ket}
\left|\nu,\dot{\nu};\lambda,\dot{\lambda};\mu,\dot{\mu};\sigma,\dot{\sigma}\right\rangle,
\end{equation}
where
$$\nu,\dot{\nu}\in\lf 0,1/2,1,3/2,\ldots\rf,$$ $$\lambda\in\lf-\nu,-\nu+1,\ldots,\nu-1,\nu\rf,\quad \dot{\lambda}\in\lf-\dot{\nu},-\dot{\nu}+1,\ldots,\dot{\nu}-1,\dot{\nu}\rf,$$ $$\mu\in\lf-\lambda,-\lambda+1,\ldots,\lambda-1,\lambda\rf,\quad \dot{\mu}\in\lf-\dot{\lambda},-\dot{\lambda}+1,\ldots,\dot{\lambda}-1,\dot{\lambda}\rf,$$ $$\sigma,\dot{\sigma}\in\lf-1/2,+1/2\rf.$$
The relationship of the numbers included in the ket vector (\ref{Ket}) with the quantum numbers $n$, $l$, $m$ and $s$ of the Madelung basis ket vector (\ref{Ket1}:
\[
\left|n,l,m,s\right\rangle
\]
(see Table \,1 in section 2) is given by the following relations
\[
n=\left|\nu-\dot{\nu}\right|,\quad l=\left|\lambda-\dot{\lambda}\right|,\quad m=\left|\mu-\dot{\mu}\right|,\quad
s=\left|\sigma-\dot{\sigma}\right|.
\]
In this case, the main quantum number $n$ varies within
\begin{equation}\label{Range}
-\left|\nu-\dot{\nu}\right|,\;-\left|\nu-\dot{\nu}\right|+1,\;-\left|\nu-\dot{\nu}\right|+2,\;\ldots,\;
\left|\nu-\dot{\nu}\right|.
\end{equation}
Figure \ref{pic_MA} shows a combined diagram of two three-dimensional projections of the weight diagram of the algebra $\mathfrak{so}(4,4)$. The upper tower, starting with hydrogen \textbf{H}, represents a \textit{pyramid of matter}, where the proper numbers of the radial generator $\bsL_{56}=\Delta_3$ take positive values. The lower tower, reflected from the plane formed by the axes of the generators $\bsL_{12}=\sL_3$ and $\bsL_{34}=\sA_3$, starting with antihydrogen $\overline{\mathbf{H}}$, represents the \textit{antimatter pyramid}, where the eigenvalues of the generator $\Delta_3$ takes negative values.

\end{document}